\documentclass[aps,prl,reprint,superscriptaddress]{revtex4-1}

\usepackage{graphicx}
\usepackage{siunitx}
\usepackage{amsmath}
\usepackage{braket}
\usepackage{xcolor}
\usepackage[colorlinks]{hyperref}

\newcommand{\nm}{\overline{n}_\mathrm{m}}
\newcommand{\gm}{\gamma_\mathrm{m}}

\begin{document}

\title{Dissipation-based Quantum Sensing of Magnons with a Superconducting Qubit}

\newcommand{\affRCAST}{Research Center for Advanced Science and Technology (RCAST), The University of Tokyo, Meguro, Tokyo 153-8904, Japan}
\newcommand{\affRIKEN}{Center for Emergent Matter Science (CEMS), RIKEN, Wako, Saitama 351-0198, Japan}
\newcommand{\affKIS}{Komaba Institute for Science (KIS), The University of Tokyo, Meguro, Tokyo 153-8902, Japan}
\newcommand{\affNQ}{Nord Quantique, Sherbrooke, Qu\'ebec, J1K 0A5, Canada}

\author{S. P. Wolski}
\email[]{swolski@qc.rcast.u-tokyo.ac.jp}
\affiliation{Research Center for Advanced Science and Technology (RCAST), The University of Tokyo, Meguro, Tokyo 153-8904, Japan}
\author{D. Lachance-Quirion}
\altaffiliation{Present address: Nord Quantique, Sherbrooke, Qu\'ebec, J1K 0A5, Canada}
\affiliation{Research Center for Advanced Science and Technology (RCAST), The University of Tokyo, Meguro, Tokyo 153-8904, Japan}
\author{Y. Tabuchi}
\affiliation{Research Center for Advanced Science and Technology (RCAST), The University of Tokyo, Meguro, Tokyo 153-8904, Japan}
\author{S. Kono}
\affiliation{Research Center for Advanced Science and Technology (RCAST), The University of Tokyo, Meguro, Tokyo 153-8904, Japan}
\affiliation{Center for Emergent Matter Science (CEMS), RIKEN, Wako, Saitama 351-0198, Japan}
\author{A. Noguchi}
\affiliation{Komaba Institute for Science (KIS), The University of Tokyo, Meguro, Tokyo 153-8902, Japan}
\author{K. Usami}
\affiliation{Research Center for Advanced Science and Technology (RCAST), The University of Tokyo, Meguro, Tokyo 153-8904, Japan}
\author{Y. Nakamura}
\email[]{yasunobu@ap.t.u-tokyo.ac.jp}
\affiliation{Research Center for Advanced Science and Technology (RCAST), The University of Tokyo, Meguro, Tokyo 153-8904, Japan}
\affiliation{Center for Emergent Matter Science (CEMS), RIKEN, Wako, Saitama 351-0198, Japan}

\date{\today}

\begin{abstract}
Hybrid quantum devices expand the tools and techniques available for quantum sensing in various fields.
Here, we experimentally demonstrate quantum sensing of the steady-state magnon population in a magnetostatic mode of a ferrimagnetic crystal.
Dispersively coupling the magnetostatic mode to a superconducting qubit allows the detection of magnons using Ramsey interferometry with a sensitivity on the order of $10^{-3}$~magnons/$\sqrt{\text{Hz}}$.
The protocol is based on dissipation as dephasing via fluctuations in the magnetostatic mode reduces the qubit coherence proportionally to the number of magnons.
\end{abstract}

\maketitle



Quantum states are intrinsically fragile with regards to external perturbations. 
This property is leveraged in quantum sensing, where appropriate quantum systems can be monitored to detect a signal~\cite{Degen2017}.
Superconducting qubits are attractive candidates for quantum sensors~\cite{Degen2017,Bylander2011,Narla2016,Inomata2016,Kono2018,Kristen2019a,Wang2019c,Honigl-Decrinis2020} as their large electric dipole moment enables strong coupling to electromagnetic fields~\cite{Blais2004, Wallraff2004}.
Recent developments of hybrid quantum systems extend the range of applicability of qubits as quantum sensors through coupling the qubits to additional degrees of freedom~\cite{Kurizki2015,Schleier-Smith2016,Clerk2020,Satzinger2018,Chu2018,Arrangoiz-Arriola2019,Lachance-Quirion2020}.

Magnons, the quanta of collective spin excitations in magnetically-ordered systems~\cite{Gurevich1996,Stancil2009}, provide a rich emerging platform for advances in quantum technologies~\cite{Tabuchi2015,Tabuchi2016,Lachance-Quirion2019,Haigh2016a,Hisatomi2016,Zhang2016,Kusminskiy2016,Wang2019}.
The presence of large quantities of magnons is typically detected using techniques such as electromagnetic induction~\cite{Gurevich1996}, the inverse spin-Hall effect~\cite{Saitoh2006,Kajiwara2010,Chumak2012}, 
or Brillouin light scattering~\cite{Demokritov2001,Sebastian2015a}.
Recently, single-shot detection of a single magnon was demonstrated in a superconducting-qubit-based hybrid quantum system, bringing the equivalent of a high-efficiency single-photon detector to the field of magnonics~\cite{Lachance-Quirion2020}. 
Such an approach, carried out by entangling the qubit and magnetostatic mode, can be used to verify that a magnon is present at a given time.
However, a different measurement scheme is desired to detect a steady-state magnon population, for example when characterizing weak continuous magnon creation processes.

In this Letter, we demonstrate dissipation-based quantum sensing of magnons in a magnetostatic mode by utilizing a transmon qubit as a quantum sensor.
The hybrid device architecture allows for an engineered dispersive interaction between the qubit and magnetostatic mode, operated in the strong dispersive regime~\cite{Gambetta2006,Schuster2007,Lachance-Quirion2017,Lachance-Quirion2019,Arrangoiz-Arriola2019,Lachance-Quirion2020}.
Fluctuations of the magnon number in the magnetostatic mode induce dephasing in the qubit in proportion to the magnon population and, as such, measurements of the coherence of the qubit yield information about the average number of magnons in the mode~\cite{Lachance-Quirion2017}.
Characterization of the sensing procedure reveals a magnon detection sensitivity on the order of~$10^{-3}$~$\mathrm{magnons}/\sqrt{\mathrm{Hz}}$, in good agreement with numerical simulations.

\begin{figure*}[t!]
  \includegraphics[]{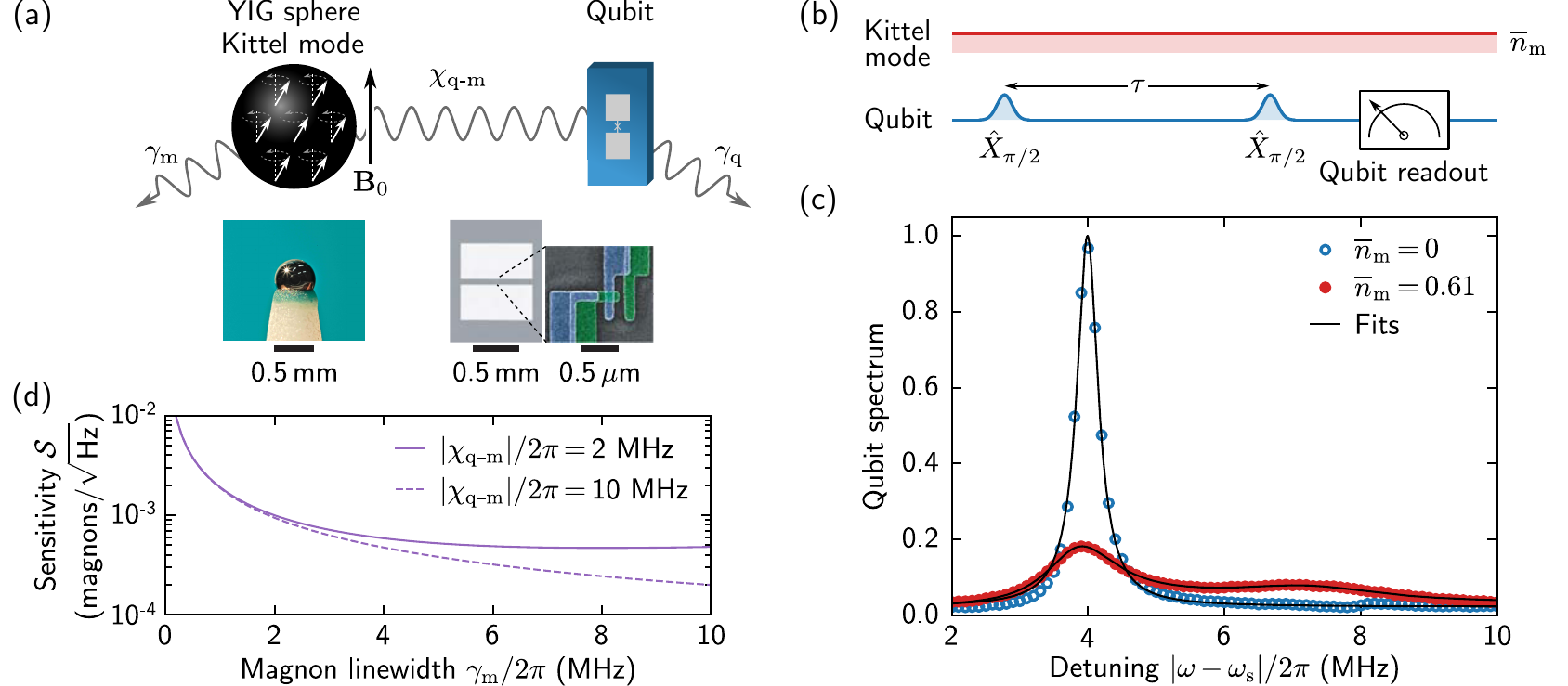}
  \caption{
    (a)~Photographs and schematics of a 0.5-mm-diameter single-crystal yttrium-iron-garnet (YIG) sphere and a superconducting transmon qubit. 
    A static magnetic field $\mathbf{B}_0$ magnetizes the YIG sphere to saturation. 
    The Kittel mode, the magnetostatic mode corresponding to uniform spin precession, couples dispersively to the qubit with a dispersive shift $\chi_\textrm{q--m}$. 
    The Kittel mode and qubit have linewidths $\gamma_\mathrm{m}$ and $\gamma_\mathrm{q}$, respectively.
    (b)~Pulse sequence used for the quantum sensing of magnons. 
    Two $\hat{X}_{\pi/2}$ pulses with frequency $\omega_\mathrm{s}$, separated by the sensing time $\tau$, are applied to the qubit as in conventional Ramsey interferometry, followed by a readout of the qubit state. A continuous drive close to resonance with the Kittel mode creates a coherent state of magnons with population~$\overline{n}_\mathrm{m}$.
    (c)~Qubit spectra obtained from Ramsey interferometry with a steady-state magnon population $\overline{n}_\mathrm{m} = 0$ (blue circles) and 0.61 (red dots). 
    The detuning $\left|\omega-\omega_\mathrm{s}\right|/2\pi$ is relative to the qubit control frequency $\omega_\mathrm{s}/2\pi$, which is itself detuned from the qubit frequency by~$\SI{-4}{\MHz}$ to induce Ramsey oscillations from which the spectra are obtained.
    Solid black lines show fits to a model.
    The spectra are normalized to the maximal value of the fit for $\overline{n}_\mathrm{m} = 0$.
    (d)~Magnon detection sensitivity~$\mathcal{S}$ calculated as a function of the Kittel mode linewidth $\gamma_\mathrm{m}/2\pi$, for amplitudes of the qubit--magnon dispersive shift \mbox{$\left|\chi_\textrm{q--m}\right|/2\pi = \SI{2}{\MHz}$} (solid line) and $\SI{10}{\MHz}$ (dashed line).
    The sensing time $\tau = \SI{0.89}{\mu\second}$ is fixed to be equal to the qubit coherence time $T_2^\ast = 2/\gamma_\mathrm{q}$ measured in the experiment.
    }
  \label{fig1}
\end{figure*}

The hybrid system used in the experiments consists of a transmon-type superconducting qubit and a single-crystal yttrium-iron-garnet (YIG) sphere~[Fig.\;\ref{fig1}(a)], both mounted inside a three-dimensional microwave copper cavity~\cite{Tabuchi2015,Tabuchi2016}.
The $\mathrm{TE}_{102}$ mode of the cavity has a dressed frequency of \mbox{$\SI{8.448}{\GHz}$}. 
The transmon qubit has a dressed frequency of \mbox{$\SI{7.914}{\GHz}$} and anharmonicity of \mbox{$\SI{-0.123}{\GHz}$}.
A magnetic circuit consisting of a pair of permanent magnets, an iron yoke, and a superconducting coil is used to apply a uniform magnetic field $\mathbf{B}_0$ to the YIG sphere, magnetizing it to saturation. 
The amplitude of the applied magnetic field, which can be tuned by changing the current in the coil, sets the frequency $\omega_\mathrm{m}$ of the uniform magnetostatic mode, or Kittel mode, in the YIG sphere~\cite{Tabuchi2014}.
The Kittel mode couples to the cavity mode through a magnetic dipole interaction of $\SI{23}{\MHz}$~\cite{Huebl2013, Tabuchi2014}.
Similarly, the qubit couples to the cavity mode through an electric dipole interaction of \mbox{$\SI{130}{\MHz}$}~\cite{Blais2004, Wallraff2004, Koch2007, Paik2011}. 
The mutual couplings with the cavity result in an effective coupling between the qubit and Kittel mode of \mbox{$\SI{7.07}{\MHz}$}~\cite{Tabuchi2015,Tabuchi2016}.

Here, the qubit and Kittel mode are detuned so that the coupling between them is dispersive, characterized by a dispersive shift per excitation, $2\chi_\textrm{q--m}$. 
At such an operating point, direct energy exchange between the qubit and Kittel mode is suppressed, however, the qubit frequency becomes dependent on the magnon state~\cite{Gambetta2006,Schuster2007,Lachance-Quirion2017}.
Measuring the qubit state thus yields different results depending on the number of magnons present in the Kittel mode. 
In the experiment, the Kittel mode frequency \mbox{$\omega_\mathrm{m}/2\pi\approx\SI{7.781}{\GHz}$} is fixed such that the system is in the strong dispersive regime, where the dispersive shift is greater than the linewidth of either system~\cite{Gambetta2006,Schuster2007,Lachance-Quirion2017}.
This can be verified via Ramsey interferometry as in Fig.\;\ref{fig1}(b), revealing the magnon-number splitting of the qubit spectrum as in Fig.\;\ref{fig1}(c)~\cite{Lachance-Quirion2019,Lachance-Quirion2020}. 
The shift per excitation is \mbox{$2\chi_\textrm{q--m}/2\pi=\SI{-3.48}{\MHz}$}, compared to the qubit linewidth~\mbox{$\gamma_\mathrm{q}/2\pi=\SI{0.36}{\MHz}$} and magnon linewidth~\mbox{$\gamma_\mathrm{m}/2\pi=\SI{1.6}{\MHz}$}.
Magnons in the Kittel mode induce both increased dephasing and a continuous frequency shift of the qubit. 
However, for the qubit resonance for the magnon Fock state $n_\mathrm{m}=0$, only the former remains in the strong dispersive regime as the latter is suppressed~\cite{Gambetta2006,SM}.

Sensing of a steady-state population of a coherent state of magnons in the Kittel mode is carried out by performing Ramsey interferometry on the qubit.
The sensitivity $\mathcal{S}$ is defined as the smallest measurable value of the magnon population that can be detected with a unit signal-to-noise ratio over a \mbox{one-second} integration time~\cite{Degen2017}.
Figure~\ref{fig1}(d) shows the magnon detection sensitivity~$\mathcal{S}$ calculated from an analytical model as a function of the magnon linewidth~$\gamma_\mathrm{m}$ for two values of the dispersive shift amplitude~$\left|\chi_\textrm{q--m}\right|$, with all other parameters similar to those in the experiment~\cite{SM}.
For values up to $\gamma_\mathrm{m}\approx4\lvert\chi_\textrm{q--m}\rvert$, increasing the magnon linewidth improves the sensitivity with a scaling $\mathcal{S}\sim1/\gm$, demonstrating that the sensing is governed primarily by dissipation in the Kittel mode.
Notably, within the strong dispersive regime with $\gamma_\mathrm{m} < 2\lvert\chi_\textrm{q--m}\rvert$, further increasing the amplitude of the dispersive shift $\lvert\chi_\textrm{q--m}\rvert$ has a negligible effect on the sensitivity as the magnon-number peaks are already sufficiently resolved.

\begin{figure}[t!]
  \includegraphics[]{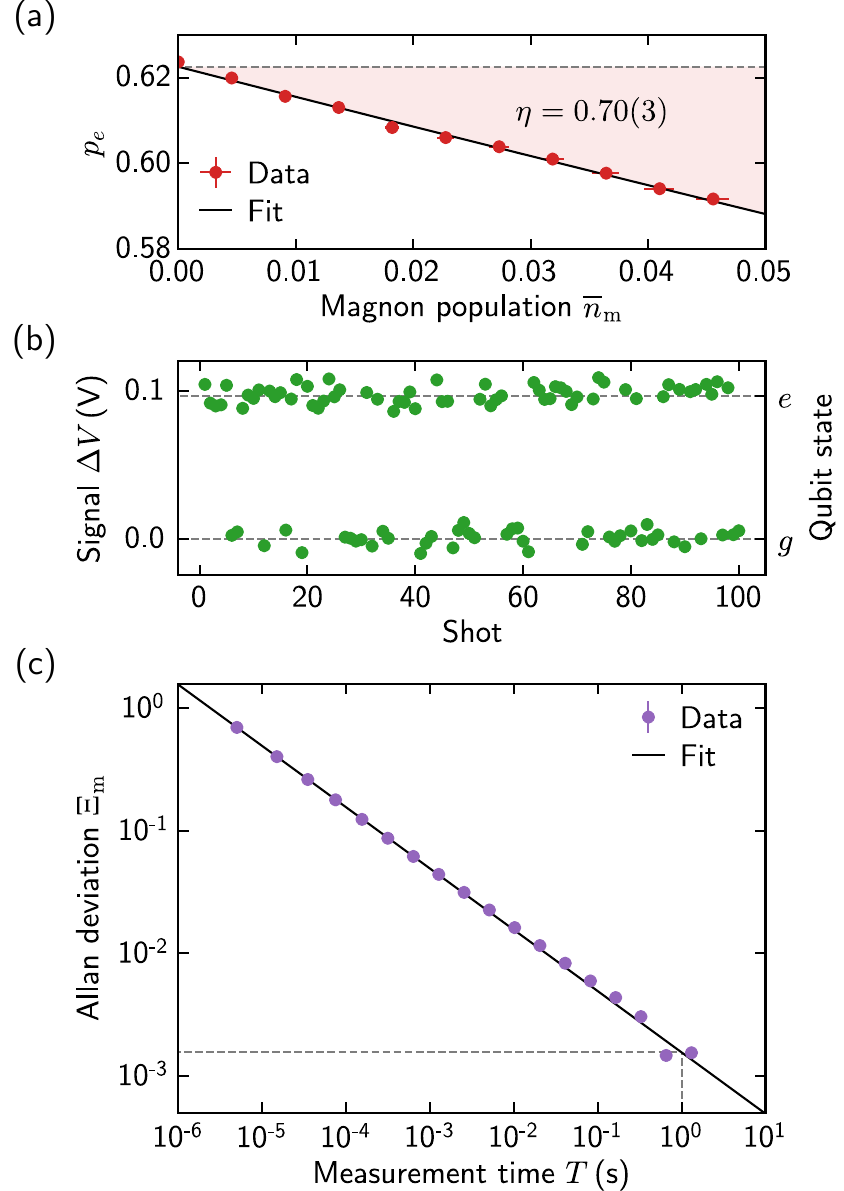}
  \caption{
    (a) Qubit excited-state probability $p_e$ as a function of the magnon population~$\overline{n}_\mathrm{m}$. 
    Equation \eqref{eq:pe-vs-nm} is fitted to the data to yield the magnon detection \mbox{efficiency~$\eta = 0.70$.}
    (b) Demodulated signal~$\Delta V$ for a subset of single-shot qubit measurements. Horizontal dashed lines indicate the demodulated signal corresponding to the qubit in the ground state~$\ket{g}$ and excited state~$\ket{e}$.
    (c) Magnon population Allan deviation $\Xi_\mathrm{m}$ as a function of measurement time~$T$. 
    The solid line indicates a fit to Eq.\;\eqref{eq:allan-dev-sens}. 
    The horizontal dashed line indicates the value of $\Xi_\mathrm{m}$ for $T=\SI{1}{\second}$ (vertical dashed line), corresponding to the magnon detection \mbox{sensitivity $\mathcal{S} = 1.55\times 10^{-3}$~$\mathrm{magnons}/\sqrt{\mathrm{Hz}}$}.
    }
  \label{fig2}
\end{figure}

The sensitivity is benchmarked by using a microwave drive applied near resonance with the Kittel mode to excite a coherent state of magnons with an average population $\overline{n}_\mathrm{m}$ with the qubit in the ground state~[Fig.\;\ref{fig1}(b)].
The magnon population excited by a given drive amplitude is calibrated from the qubit spectrum as in Fig.\;\ref{fig1}(c)~\cite{SM}.
During the continuous magnon excitation, two $\pi/2$ pulses around the same axis are applied to the qubit, separated by a free evolution time corresponding to the sensing time $\tau$.
At the end of the sequence, the qubit state is measured using the high-power readout technique~\cite{Reed2010}. 
Due to the dispersive interaction, the probability of the qubit being in the excited state $p_e$ depends on the magnon population $\overline{n}_\mathrm{m}$, as in Fig.\;\ref{fig2}(a) for $\tau=\SI{0.8}{\mu\second}$ and qubit drive detuning $\Delta_\mathrm{s}=0$.
For~$\overline{n}_\mathrm{m}\ll 1$, this dependence can be approximated with
\begin{equation}\label{eq:pe-vs-nm}
  p_e(\overline{n}_\mathrm{m}) = p_e(0) \pm \eta\overline{n}_\mathrm{m},
\end{equation}
where the sign depends on the qubit drive detuning and sensing time, and $\eta$ is the efficiency of detecting magnons with the qubit, defined to be positive~\cite{SM}. 
A fitted value of \mbox{$\eta = 0.70(3)$} is obtained from the data of Fig.\;\ref{fig2}(a). 
The sensing signal corresponds to the difference between the probability $p_e$ with and without magnons present in the Kittel mode.
Indeed, the qubit excitation probability $p_e$ can be expressed in terms of the magnon population $\overline{n}_\mathrm{m}$ through Eq.\;\eqref{eq:pe-vs-nm}, using the previously obtained value of $\eta$.

To characterize the noise associated with the sensing procedure, repeated single-shot readout of the qubit, shown in Fig.\;\ref{fig2}(b), is carried out using the high-power readout technique~\cite{Reed2010}. 
Such a measurement also allows for quantifying the fidelity of the qubit readout, here estimated at around $90\%$~\cite{SM}.
The scaling of sensor noise for different measurement times is given by the Allan deviation as shown in Fig.\;\ref{fig2}(c), which can be expressed in terms of the qubit probability,~$\Xi_\mathrm{q}(T)$, or the magnon population,~\mbox{$\Xi_\mathrm{m}(T)=\Xi_\mathrm{q}(T)/\eta$}. 
This corresponds to the standard deviation of subsets of the data as a function of measurement \mbox{time $T = N\tau_\mathrm{total}$}, where $N$ is the number of shots and $\tau_\mathrm{total} = \SI{5}{\mu\second}$ is the duration of a single sequence.
As shown in Fig.\;\ref{fig2}(c), the sensitivity~$\mathcal{S}$ is related, by definition, to the Allan deviation via
\begin{equation}\label{eq:allan-dev-sens}
  \Xi_\mathrm{m}(T) = \mathcal{S}/\sqrt{T}.
\end{equation}
Fitting Eq.\;\eqref{eq:allan-dev-sens} to the data yields a sensitivity of \mbox{$\mathcal{S} = 1.55(5)\times10^{-3}\,\mathrm{magnons}/\sqrt{\mathrm{Hz}}$}.
The data is seen not to deviate significantly from the scaling up to $T>\SI{1}{\second}$, implying that the noise floor due to slow fluctuations has not been reached and scaling the noise to $T=\SI{1}{\second}$ is valid.
Numerical simulations modeling the dispersive qubit--magnon interaction and imitating the experimental protocol yield a sensitivity of \mbox{$1.35\times10^{-3}\,\mathrm{magnons}/\sqrt{\mathrm{Hz}}$}, showing excellent agreement with the experimental results.
A single fitting parameter is used, but it is seen not to affect the value of the sensitivity significantly at zero detuning~\cite{SM}. 
The underestimation of the sensitivity by the simulation is probably due to additional interactions that are not accounted for in the dispersive Hamiltonian considered in the simulations. 
For example, excluding the second excited state of the transmon qubit in the simulations decreases the sensitivity by up to 10\%.

\begin{figure}[t!]
  \includegraphics[]{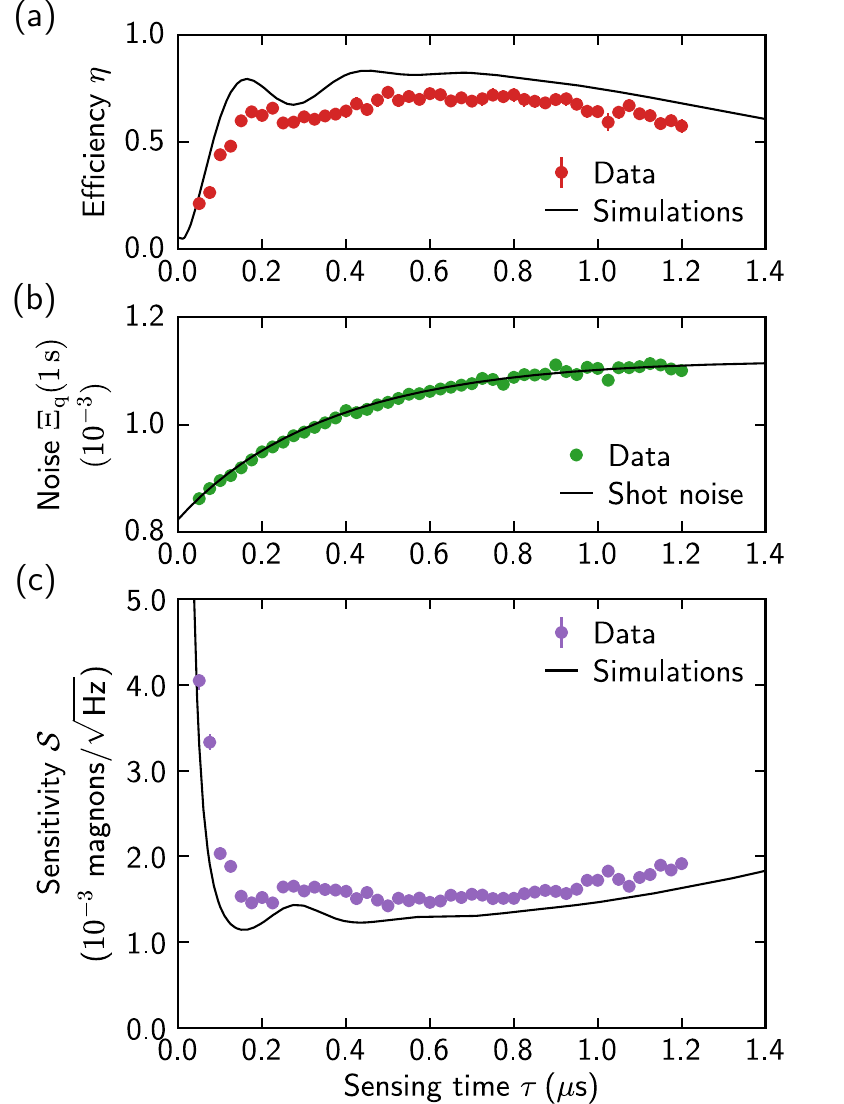}
  \caption{
    (a) Magnon detection efficiency $\eta$, 
    (b)~noise of single-shot qubit readout for one-second measurement time,~\mbox{$\Xi_\mathrm{q}(\SI{1}{\second})$}, and~(c)~magnon detection sensitivity~$\mathcal{S}$ as a function of sensing time~$\tau$.
    In (a) and~(c), solid black lines are results from numerical simulations.
    In (b), the solid black line is the shot noise calculated from Eq.\;\eqref{eq:shot-noise}.
    }
  \label{fig3}
\end{figure}

\begin{figure}[t!]
  \includegraphics[]{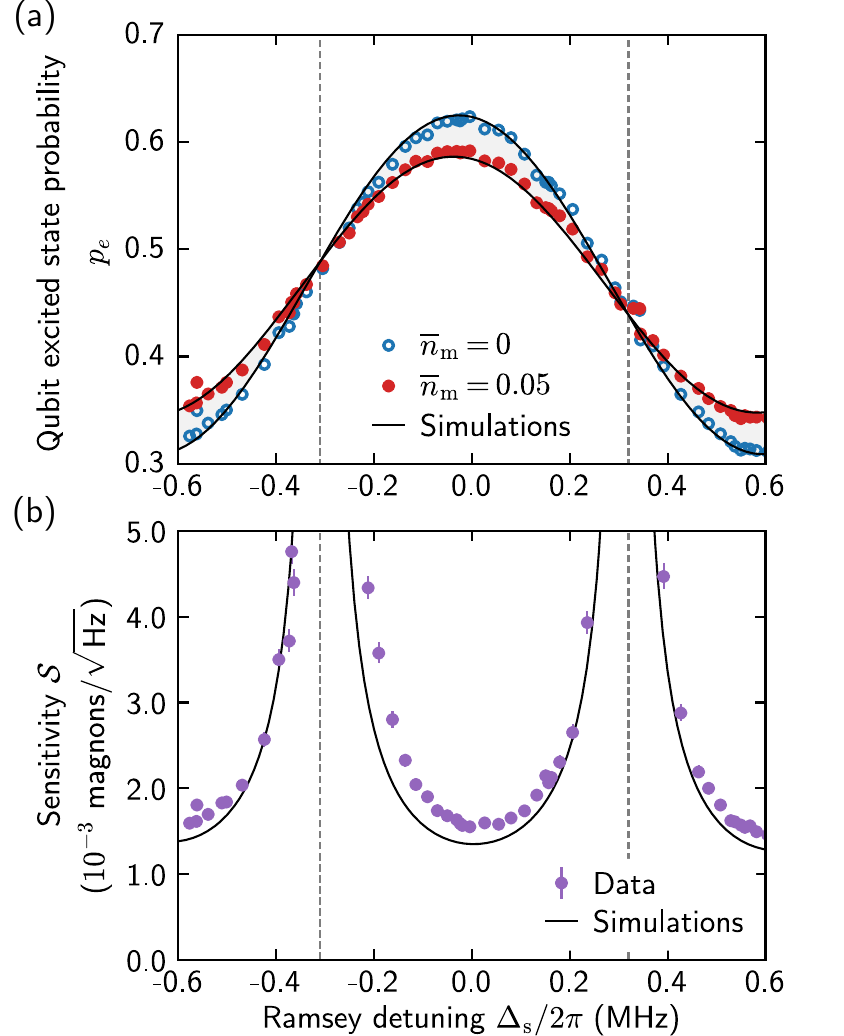}
  \caption{
    (a) Qubit excited-state probability $p_e$ and (b)~magnon detection sensitivity~$\mathcal{S}$ as a function of Ramsey detuning~$\Delta_\mathrm{s}/2\pi$. 
    The probability $p_e$ is shown for the cases of sensing different steady-state magnon populations $\overline{n}_\mathrm{m}=0$ (blue circles) and $\overline{n}_\mathrm{m}=0.05$ (red dots), with the shaded region indicating the difference. 
    Dashed vertical lines correspond to nodes of the Ramsey fringes. 
    Solid black lines are results from numerical simulations.
    }
  \label{fig4}
\end{figure}


The sensing time $\tau$, here corresponding to the free evolution time between the $\hat{X}_{\pi/2}$ pulses in the Ramsey sequence, is a fundamental parameter that can be used to investigate the performance of the sensor.
A longer sensing time increases the acquisition of information, improving sensitivity. 
However, a longer sensing time also leads to an increased loss of qubit coherence, resulting in a worse sensitivity; this implies a trade-off for an optimal sensing time. 
The efficiency~$\eta$ is measured as a function of sensing time~$\tau$, as in Fig.\;\ref{fig3}(a) for $\Delta_\mathrm{s}=0$. 
This reveals a weak optimum, as  information acquisition saturates rapidly but qubit decoherence has a relatively slower falloff with $T_2^\ast = \SI{0.89}{\mu s}$.
Numerical simulations, the results of which are shown as a solid line in Fig.\;\ref{fig3}(a), are in good agreement with the data and reproduce all essential features. 
The oscillations are due to dynamical detuning of the magnon frequency as the excited state of the qubit is populated during the sensing sequence.

Characterization of the noise associated with repeated measurements is also carried out as a function of sensing time, calculating the Allan deviation of the qubit excitation probability for a one-second measurement time,~\mbox{$\Xi_\mathrm{q}(T=\SI{1}{\second})$}, as shown in Fig.\;\ref{fig3}(b) for $\Delta_\mathrm{s}=0$.
The solid line in Fig.\;\ref{fig3}(b) shows the qubit shot noise \cite{Degen2017} given by
\begin{equation}\label{eq:shot-noise}
  \Xi_\mathrm{q}(T=\SI{1}{\second}) = \sqrt{ p_e\left(1-p_e\right)\frac{\tau_\mathrm{total}}{\SI{1}{\second}}},
\end{equation}
where $p_e = p_e(\nm=0)$.
This shows excellent agreement between theory and experiment, as the qubit readout is shot-noise limited due to the high readout fidelity.
In other words, the high-fidelity projective qubit readout ensures that the qubit readout process is not the dominant constraint on the magnon detection sensitivity, as noise from components such as amplifiers has been digitized.
Figure \ref{fig3}(c) shows the sensitivity $\mathcal{S}$ as a function of the sensing time $\tau$ obtained from the efficiency~$\eta$ and noise~$\Xi_\mathrm{q}$.
The optimal sensing time is seen to be approximately half of the qubit coherence time \mbox{$T_2^\ast = \SI{0.89}{\mu\s}$}, with the best sensitivity \mbox{$\mathcal{S} = 1.42(3)\times10^{-3}\,\mathrm{magnons}/\sqrt{\mathrm{Hz}}$} measured for \mbox{$\tau = \SI{0.5}{\mu\second}$}.
This is around twice as fast as procedures based on sensing via a frequency shift, where the optimal sensing time is equal to the coherence time~\cite{Degen2017}.


The relative contributions of dephasing and frequency shifts to the sensitivity are revealed in this case by examining the Ramsey fringes as in Fig.\;\ref{fig4}(a).
Here, these are measured by sweeping the relative detuning $\Delta_\mathrm{s}$ between the frequencies of the qubit and the~$\pi/2$ pulses applied to the qubit during the sensing protocol.
The Ramsey fringes show a reduction of contrast when a finite magnon population is present in the Kittel mode, but do not exhibit a significant frequency shift.
Figure \ref{fig4}(b) shows that the optimal sensitivity is obtained near zero detuning, and becomes asymptotically worse when approaching detunings corresponding to nodes of the Ramsey fringes.
Consequently, the dissipation of magnons is verified as the dominant mechanism by which the qubit is sensitive to the magnon population within the strong dispersive regime, and increasing the linewidth of the Kittel mode up to \mbox{$\gamma_\mathrm{m}\approx4\lvert\chi_\textrm{q--m}\rvert$} would improve the sensitivity by as much as a factor of three without other changes~\cite{SM}.

These conclusions reflect the qualitative behaviour explored in Fig.\;\ref{fig1}(d), and thus the procedure presented here falls under the category of sensing techniques known as $T_2^\ast$ relaxometry~\cite{Degen2017}.
Such a conclusion contrasts with previous approaches such as in Ref.\;\cite{Lachance-Quirion2020}, where the sensing of magnons is based on entanglement of the qubit and Kittel mode, and is thus greatly hindered by an increased magnon dissipation rate, even in the strong dispersive regime.


The demonstrated level of sensitivity represents a significant advancement relative to existing magnon detection schemes, where the typical quantity of detected magnons is many orders of magnitude larger~\cite{Kajiwara2010,Chumak2012,Cramer2018}.
Furthermore, detection on the level of single magnons and below can be useful for probing magneto-optical effects in the quantum regime as part of the development of quantum transducers~\cite{Hisatomi2016,Lachance-Quirion2019}, and may be used in dark matter searches for axion-like particles~\cite{Barbieri2017,Crescini2018, Flower2019, Crescini2020}.
The device can also be used as a static magnetic field sensor, as the detuning of the Kittel mode by such a field is measurable by monitoring the magnon population excited by a fixed microwave drive.
In this way, the magnetic-field-insensitive transmon qubit can be made sensitive to the applied field in a controlled manner. 

The Kittel mode linewidth can be increased by doping the ferrimagnetic crystal with rare-earth ions~\cite{Sparks1964}, or increasing the radiative decay of the Kittel mode~\cite{Yao2019}.
Additionally, the qubit lifetime is Purcell-limited, so the coherence time $T_2^\ast$ of the qubit can be improved by reducing cavity losses, allowing a longer sensing time with the same degree of coherence.
Interactions beyond the dispersive qubit--magnon interaction, such as that due to the cavity--magnon cross-Kerr interaction~\cite{Helmer2009,Lachance-Quirion2019}, could also be utilized for magnon sensing.
They could potentially offer further improvements in sensitivity alongside enabling continuous sensing.

In conclusion, we have demonstrated that probing the coherence of a superconducting transmon qubit dispersively coupled to a magnetostatic mode allows for quantum sensing of magnons with a sensitivity on the order of $10^{-3}$~$\mathrm{magnons}/\sqrt{\mathrm{Hz}}$.
The device parameters and operating point of the protocol lead to the qubit becoming sensitive to the magnon population primarily through magnon decay.
Counterintuitively, this results in the detection sensitivity being improved by increasing losses in the magnetostatic mode while operating close to or in the strong dispersive regime.
The results presented here constitute an advancement in the detection and characterization of small magnon populations, and provide a tool for fields such as magnonics, magnon spintronics, and hybrid quantum systems, as well as fundamental science of magnons and related phenomena such as dark matter detection.

\begin{acknowledgments}
The authors would like to thank Arjan van Loo for valuable discussions, as well as the photograph of the YIG sphere.
This work was partly supported by JSPS KAKENHI (26220601, 18F18015), JST ERATO (JPMJER1601), and FRQNT Postdoctoral Fellowships. S.P.W.~was supported by the MEXT Monbukagakusho Scholarship. D.L.-Q.~was an International Research Fellow of JSPS.
\end{acknowledgments}

\bibliographystyle{apsrev4-1}
%

\end{document}


\title{Supplementary Materials for ``Dissipation-based Quantum Sensing of Magnons with a Superconducting Qubit''}

\newcommand{\affRCAST}{Research Center for Advanced Science and Technology (RCAST), The University of Tokyo, Meguro, Tokyo 153-8904, Japan}
\newcommand{\affRIKEN}{Center for Emergent Matter Science (CEMS), RIKEN, Wako, Saitama 351-0198, Japan}
\newcommand{\affKIS}{Komaba Institute for Science (KIS), The University of Tokyo, Meguro, Tokyo 153-8902, Japan}
\newcommand{\affNQ}{Nord Quantique, Sherbrooke, Qu\'ebec, J1K 0A5, Canada}

\author{S. P. Wolski}
\email[]{swolski@qc.rcast.u-tokyo.ac.jp}
\affiliation{\affRCAST}
\author{D. Lachance-Quirion}
\altaffiliation{Present address: \affNQ}
\affiliation{\affRCAST}
\author{Y.~Tabuchi}
\affiliation{\affRCAST}
\author{S.~Kono}
\affiliation{\affRCAST}
\affiliation{\affRIKEN}
\author{A.~Noguchi}
\affiliation{\affKIS}
\author{K.~Usami}
\affiliation{\affRCAST}
\author{Y.~Nakamura}
\email[]{yasunobu@ap.t.u-tokyo.ac.jp}
\affiliation{\affRCAST}
\affiliation{\affRIKEN}

\date{\today}

\maketitle
\tableofcontents
\newpage

\section{Theory}\label{sec:theory}

\subsection{Governing equations}\label{sec:theory-governing-equations}

The $\mathrm{TE}_{10p}$ cavity modes, labeled by the index $p$, are described as harmonic oscillators with
\begin{equation}
  \hat{\mathcal{H}}_\mathrm{c}/\hbar = \sum_p \omega_\cavp \hat{a}_p^\dag \hat{a}_p,
\end{equation}
where $\hat{a}_p$ ($\hat{a}_p^\dag$) is the annihilation (creation) operator for cavity mode $\mathrm{TE}_{10p}$ with frequency $\omega_\cavp$.
The transmon qubit is described as an anharmonic oscillator with
\begin{equation}
  \hat{\mathcal{H}}_\mathrm{q}/\hbar 
  = \left( \omega_\mathrm{q} - \frac{\alpha}{2} \right) \hat{b}^\dag \hat{b}
  + \frac{\alpha}{2}\left(\hat{b}^\dag\hat{b}\right)^2,
\end{equation}
where $\hat{b}$ ($\hat{b}^\dag$) is the annihilation (creation) operator for qubit excitations, and $\omega_\mathrm{q}$ defined to be the transition frequency between the ground state $\ket{g}$ and the first excited state $\ket{e}$ of the qubit. The anharmonicity is defined such that $\omega_\mathrm{q} + \alpha$ is the transition frequency between the first excited state $\ket{e}$ and the second excited state $\ket{f}$ of the qubit. 
The Kittel mode is also described as a harmonic oscillator with
\begin{equation}
  \hat{\mathcal{H}}_\mathrm{m}/\hbar = \omega_\mathrm{m}\hat{c}^\dag\hat{c},
\end{equation}
where $\hat{c}$ ($\hat{c}^\dag$) is the annihilation (creation) operator for magnons in the Kittel mode with frequency $\omega_\mathrm{m}$. The assumption of harmonicity is motivated by the fact that the magnon population $\overline{n}_\mathrm{m}$ is much smaller than the $\sim \num{1.4e18}$ spins in the YIG sphere. Higher-order magnetostatic modes are neglected, as the uniformity of both the external magnetic field and the microwave magnetic field associated with the lowest-order cavity modes suppresses their coupling to the cavity modes and, by extension, the qubit~\cite{Tabuchi2016,Lachance-Quirion2019}.

The cavity--qubit interaction occurs through an electric dipole coupling. This can be described by the Jaynes--Cummings Hamiltonian
\begin{equation}
  \hat{\mathcal{H}}_\textrm{q--c}/\hbar = \sum_p g_{\textrm{q--}\cavp}\left(\hat{a}_p^\dag\hat{b} + \hat{a}_p\hat{b}^\dag\right),
\end{equation}
with $g_{\textrm{q--}\cavp}$ the coupling strength between the first qubit transition and $\mathrm{TE}_{10p}$ cavity mode~\cite{Blais2004}. 
Similarly, the cavity--magnon interaction occurs through a magnetic dipole coupling, with the corresponding Hamiltonian
\begin{equation}
  \hat{\mathcal{H}}_\textrm{m--c}/\hbar = \sum_p g_{\textrm{m--}\cavp}\left(\hat{a}_p^\dag\hat{c} + \hat{a}_p\hat{c}^\dag\right),
\end{equation}
with $g_{\textrm{m--}\cavp}$ the coupling strength between the Kittel mode and the $\mathrm{TE}_{10p}$ cavity mode~\cite{Tabuchi2016,Lachance-Quirion2019}.

The complete hybrid system can thus be described by the Hamiltonian
\begin{equation}
  \hat{\mathcal{H}} = \hat{\mathcal{H}}_\mathrm{c} + \hat{\mathcal{H}}_\mathrm{q} + \hat{\mathcal{H}}_\mathrm{m} + \hat{\mathcal{H}}_\textrm{q--c} + \hat{\mathcal{H}}_\textrm{m--c},
\end{equation}
which is numerically diagonalized to obtain the qubit--magnon coupling strength~$g_\textrm{q--m}$ and dispersive shift~$\chi_\textrm{q--m}$.
The cavity modes can be adiabatically eliminated when $\left|\omega_p - \omega_\mathrm{q}\right|, \left|\omega_p - \omega_\mathrm{m}\right| \gg g_{\textrm{q--}\cavp}, g_{\textrm{m--}\cavp}$, that is, the qubit and Kittel mode are far detuned from the cavity modes \cite{Tabuchi2016}.
Additionally, when $\left|\omega_\mathrm{q} - \omega_\mathrm{m}\right| \ll g_{\textrm{q--}\cavp}, g_{\textrm{m--}\cavp}$, so that the qubit and Kittel mode are relatively close to resonance, the interaction between them can be described by
\begin{equation}
  \hat{\mathcal{H}}_\textrm{q--m}/\hbar = g_\textrm{q--m}\left(\hat{b}^\dag\hat{c} + \hat{b}\hat{c}^\dag\right),
\end{equation}
with $g_\textrm{q--m}$ the coupling strength between the qubit and Kittel mode \cite{Tabuchi2015,Tabuchi2016,Lachance-Quirion2019}.
When the qubit and Kittel mode are resonant such that $\omega_\mathrm{q} = \omega_\mathrm{m} = \omega_\mathrm{q,m}$, the qubit--magnon coupling is approximately~\cite{Lachance-Quirion2020}
\begin{equation}\label{eq:g-q-m}
  g_\textrm{q--m} \approx \sum_p \frac{g_{\textrm{q--}\cavp}g_{\textrm{m--}\cavp}}{\omega_\cavp-\omega_\mathrm{q,m}}.
\end{equation}

\begin{table*}[t]
  \caption{Cavity parameters. Values in square brackets indicate the parameters have not been measured directly, but have been evaluated based on other measured parameters.}
  \begin{ruledtabular}
  \begin{tabular*}{\textwidth}{ll@{\extracolsep{\fill}}d{5}d{5}d{5}d{5}}
    Parameter & Unit & \multicolumn{4}{c}{Value} \\
    \hline
    Index $p$ for cavity mode $\mathrm{TE}_{10p}$ && \multicolumn{1}{c}{1} & \multicolumn{1}{c}{2} & \multicolumn{1}{c}{3} & \multicolumn{1}{c}{4} \\
    \hline
    Bare frequency $\omega_\cavp/2\pi$ & GHz & 6.989\,83 & 8.411\,63 & 10.438\,65 & [12.920\,2] \\
    Dressed frequency $\omega_\cavp^g/2\pi$ &GHz & 6.982\,70 & 8.448\,35 & 10.445\,98 & [12.922\,9] \\
    \hline
    Total linewidth $\kappa_\cavp/2\pi$ &MHz & 1.28 & 1.99 & 3.19 & -\\
    Input coupling rate $\kappa_\cavp^\mathrm{in}/2\pi$ &MHz & 0.33 & 0.73 & 0.23 & -\\
    Output coupling rate $\kappa_\cavp^\mathrm{out}/2\pi$ &MHz & 0.13 & 0.51 & 1.27 & -\\
    Internal losses $\kappa_\cavp^\mathrm{int}/2\pi$ &MHz & 0.82 & 0.76 & 1.68 & -\\
    \hline
    Electric dipole coupling to qubit $g_{\textrm{q--}\cavp}/2\pi$ &MHz & [83.2] & [128.9] & [134.8] & [116.5] \\
    Magnetic dipole coupling to Kittel mode $g_{\textrm{m--}\cavp}/2\pi$ &MHz & [-15.37] & 23.0 & [-21.7] & [12.83] \\
  \end{tabular*}
\end{ruledtabular}
  \label{tab:params-cavity}
\end{table*}

\begin{table*}[t]
  \caption{Qubit and Kittel mode parameters. Values in square brackets indicate the parameters have not been measured directly, but have been evaluated based on other measured parameters. The coil currents $I=\SI{-8.10}{\milli\ampere}$ and $\SI{6.31}{\milli\ampere}$ correspond to the Kittel mode in the strong dispersive regime with the qubit and on resonance with the $\mathrm{TE}_{102}$ cavity mode, respectively.}
  \begin{ruledtabular}
  \begin{tabular*}{\textwidth}{ll@{\extracolsep{\fill}}d{6}}
    Parameter &Unit & \multicolumn{1}{c}{Value} \\
    \hline
    Bare qubit $g$--$e$ transition frequency $\omega_\mathrm{q}/2\pi$ &GHz & [7.959\,02] \\
    Bare qubit anharmonicity $\alpha/2\pi$ &GHz & [-0.143] \\
    Dressed qubit $g$--$e$ transition frequency $\omega_\mathrm{q}^0/2\pi$ &GHz & 7.922\,10 \\
    Dressed qubit anharmonicity $\alpha_0/2\pi$ &GHz & -0.123 \\
    Qubit linewidth $\gamma_\mathrm{q}^0/2\pi$ &MHz & 0.36 \\
    \hline
    Dressed magnon frequency (at $I=\SI{-8.10}{\milli\ampere}$) $\omega_\mathrm{m}^g/2\pi$ &GHz & 7.781\,08 \\
    Magnon linewidth (at $I=\SI{-8.10}{\milli\ampere}$) $\gamma_\mathrm{m}/2\pi$ &MHz & 1.567 \\
    Magnon linewidth (at $I=\SI{6.31}{\milli\ampere}$) $\gamma_\mathrm{m}/2\pi$ &MHz & 1.483 \\
    \hline
    Qubit-magnon coupling strength $g_\textrm{q--m}/2\pi$ &MHz & 7.07 \\
  \end{tabular*}
  \end{ruledtabular}
  \label{tab:params-other}
\end{table*}

\subsection{Strong dispersive regime of the qubit--magnon interaction}\label{sec:theory-strong-dispersive-regime}

The dispersive regime is reached when the detuning between the qubit and Kittel mode is much greater in amplitude than the coupling strength $g_\textrm{q--m}$ \cite{Tabuchi2015}.
In such a regime, the system can be described by the dispersive qubit--magnon Hamiltonian
\begin{equation}\label{eq:dispersive-hamiltonian}
  \mathcal{H}_\textrm{q--m}^\textrm{disp}/\hbar = 2\chi_\textrm{q--m}\hat{b}^\dag\hat{b}\hat{c}^\dag\hat{c}.
\end{equation}
The dispersive coupling $\chi_\textrm{q--m}$ is approximately given by \cite{Koch2007}
\begin{equation}
  \chi_\textrm{q--m} \approx \frac{\alpha_0 g_\textrm{q--m}^2}{\Delta_\textrm{q--m}(\Delta_\textrm{q--m}+\alpha_0)},
\end{equation}
with $\alpha_0$ the dressed qubit anharmonicity.

From the form of the dispersive Hamiltonian in Eq.\;\eqref{eq:dispersive-hamiltonian}, it is clear that direct energy exchange between the qubit and Kittel mode is suppressed. However, the qubit spectrum becomes dependent on the magnon population, and vice versa.
In particular, the qubit spectrum in the presence of magnons in the Kittel mode can be described using a model from Refs.\;\cite{Lachance-Quirion2017,Lachance-Quirion2019,Lachance-Quirion2020} based on Ref.\;\cite{Gambetta2006}.
In this model, the qubit spectrum $s_{n_\mathrm{m}}(\Delta\omega)$ for the magnon Fock state $n_\mathrm{m}$ as a function of the detuning from the qubit drive frequency~$\Delta\omega=\omega-\omega_\mathrm{s}$ is given by
\begin{equation}\label{eq:qubit-spectrum-model}
  s_{n_\mathrm{m}}(\Delta\omega)=\frac{1}{\pi}\frac{1}{n_\mathrm{m}!}\mathrm{Re}\left[\frac{\left(-\mathcal{J}\right)^{n_\mathrm{m}}e^\mathcal{J}}{\gamma_\mathrm{q}^{n_\mathrm{m}}/2-i\left(\Delta\omega-\Delta_\mathrm{s}^{n_\mathrm{m}}\right)}\right],
\end{equation}
where
\begin{align}
\mathcal{J}&=\overline{n}_\mathrm{m}^g\mathcal{A}\left(1+\mathcal{B}\right)\left(\frac{\gamma_\mathrm{m}/2-i\left(2\chi_\mathrm{q-m}+\Delta_\mathrm{d}\right)}{\gamma_\mathrm{m}/2+i\left(2\chi_\mathrm{q-m}+\Delta_\mathrm{d}\right)}\right),\\
\omega_\mathrm{q}^{n_\mathrm{m}}&=\omega_\mathrm{q}+n_\mathrm{m}\left(2\chi_\mathrm{q-m}+\Delta_\mathrm{d}\right)+\overline{n}_\mathrm{m}^g\mathcal{C}\chi_\textrm{q--m},\label{eq:qubit-freq-magnon-number}\\
\Delta_\mathrm{s}^{n_\mathrm{m}}&=\omega_\mathrm{q}^{n_\mathrm{m}}-\omega_\mathrm{s},\label{eq:qubit-drive-detuning-magnon-number}\\
\gamma_\mathrm{q}^{n_\mathrm{m}}&=\gamma_\mathrm{q}+n_\mathrm{m}\gamma_\mathrm{m}+\overline{n}_\mathrm{m}^g\mathcal{D}\gamma_\mathrm{m},\label{eq:qubit-lw-magnon-number}\\
\mathcal{C}&=(1-\mathcal{A})(1+\mathcal{B}),\\
  \mathcal{D}&=\mathcal{A}(1+\mathcal{B}),\\
  \mathcal{A}&=\frac{2\chi_\textrm{q--m}^2}{\left(\gamma_\mathrm{m}/2\right)^2+\chi_\textrm{q--m}^2+\left(\chi_\textrm{q--m}+\Delta_\mathrm{d}\right)^2},\\
  \mathcal{B}&=\frac{\left(\gamma_\mathrm{m}/2\right)^2+\Delta_\mathrm{d}^2}{\left(\gamma_\mathrm{m}/2\right)^2+\left(\Delta_\mathrm{d}+2\chi_\textrm{q--m}\right)^2},\\
  \overline{n}_\mathrm{m}^g&=\frac{\Omega_\mathrm{d}^2}{\left(\gamma_\mathrm{m}/2\right)^2+\Delta_\mathrm{d}^2},\\
  \overline{n}_\mathrm{m}^e&=\frac{\Omega_\mathrm{d}^2}{\left(\gamma_\mathrm{m}/2\right)^2+(\Delta_\mathrm{d} + 2\chi_\textrm{q--m})^2}=\mathcal{B}\overline{n}_\mathrm{m}^g. \label{eq:n-m-e-definition}
\end{align}
In the above equations, $\Delta_\mathrm{d}\equiv\omega_\mathrm{m}^g - \omega_\mathrm{d}$ is the detuning between the magnon drive and the Kittel mode frequency when the qubit is in the ground state, $\Omega_\mathrm{d}$ is the magnon excitation strength due to the magnon drive, $n_\mathrm{m}$ is the number of magnons corresponding to a specific Fock state $\ket{n_\mathrm{m}}$, $\omega_\mathrm{q}^{n_\mathrm{m}}$ and~$\gamma_\mathrm{q}^{n_\mathrm{m}}$ are respectively the frequency and linewidth of the qubit with the Kittel mode in the Fock state~$|n_\mathrm{m}\rangle$, and $\overline{n}_\mathrm{m}^g$ ($\overline{n}_\mathrm{m}^e$) is the average magnon population with the qubit in the ground state (excited state). 

The aim is to detect the magnon population present before the protocol starts, at which point the qubit is in the ground state. 
As such, where it is unambiguous, the main text and Supplementary Material use the simplified notation $\nm=\nm^g$ and $\Delta_\mathrm{s} = \Delta_\mathrm{s}^0$.

When the coupling between the qubit and Kittel mode is strong enough, the strong dispersive regime can be entered for $2\left|\chi_\text{q--m}\right|\gg\gamma_\mathrm{q},\gamma_\mathrm{m}$. In this case, the discrete ac Stark shift $n_\mathrm{m}\left(2\chi_\mathrm{q-m}+\Delta_\mathrm{d}\right)$ in Eq.\;\eqref{eq:qubit-freq-magnon-number} allows resolution of magnon Fock states in qubit spectrum as in Fig.\;1(c)~\cite{Gambetta2006,Schuster2007}.
Furthermore, the continuous ac-Stark-shift term $\overline{n}_\mathrm{m}^g\mathcal{C}\chi_\textrm{q--m}$ in Eq.\;\eqref{eq:qubit-freq-magnon-number} vanishes in the strong dispersive regime as $\mathcal{C}\to0$. 
However, the dephasing term $\overline{n}_\mathrm{m}^g\mathcal{D}\gamma_\mathrm{m}$ in Eq.\;\eqref{eq:qubit-freq-magnon-number} remains as $\mathcal{D}\to1$. 
As a result, even with the condition $\overline{n}_\mathrm{m}^g\ll1$, there is an additional contribution, proportional to $\overline{n}_\mathrm{m}^g$ and $\gamma_\mathrm{m}$, to the dephasing of the qubit peak corresponding to the zero-magnon Fock state $n_\mathrm{m}=0$.

\subsection{Qubit state dependence on magnon population}

When carrying out Ramsey interferometry of the qubit at frequency $\omega_\mathrm{q}^0$ probed with a drive at frequency $\omega_\mathrm{s}$, the excited state probability $p_e$ for a sensing time $\tau$ is given by
\begin{equation}\label{eq:pe-ramsey-plain}
  p_e(\tau) = \frac{1}{2}\left[1+\cos\left(\tau\Delta_\mathrm{s}^0\right)\exp\left(-\tau/T_2^\ast\right)\right],
\end{equation}
where $\Delta_\mathrm{s}^0\equiv\omega_\mathrm{q}^0-\omega_\mathrm{s}$ is the qubit drive detuning and $T_2^\ast = 2/\gamma_\mathrm{q}^{0}$ is the qubit coherence time~\cite{Ramsey1950}.
The effect of the magnon population on the detuning and coherence time can be accounted for by considering the expressions for $\Delta_\mathrm{s}^0$ from Eqs.\;\eqref{eq:qubit-freq-magnon-number} and \eqref{eq:qubit-drive-detuning-magnon-number} and $\gamma_\mathrm{q}^0$ from Eq.\;\eqref{eq:qubit-lw-magnon-number}. 
For small magnon populations $\nm^g\ll1$, expansion of Eq.\;\eqref{eq:pe-ramsey-plain} to first order in $\nm^g$ gives
\begin{equation}\label{eq:p-e-magnon-number}
    p_e(\nm^g) = 
    \frac{1}{2}\left[1+\cos\left(\Delta_\mathrm{s}^0\tau\right)e^{-\tau\gamma_\mathrm{q}^0/2}\right]
    - \frac{1}{2}e^{-\tau\gamma_\mathrm{q}^0/2}\left[\frac{\tau}{2}\cos\left(\tau\Delta_\mathrm{s}^0\right)\mathcal{D}\gamma_\mathrm{m}+\tau\sin\left(\tau\Delta_\mathrm{s}^0\right)\mathcal{C}\chi_\textrm{q--m}\right]\nm^g. \\
\end{equation}
In accordance with Eq.\;(1) of the main text,
\begin{equation}\label{eq:p-e-n-m-linear}
  p_e(\nm^g) = p_e(0) \pm \eta\nm^g,
\end{equation}
the efficiency $\eta$ can therefore be defined as
\begin{equation}\label{eq:eta-full}
  \eta = \mp \frac{1}{2}\tau e^{-\gamma_\mathrm{q}^0\tau/2}
  \biggl[
    \underbrace{\cos\left(\tau\Delta_\mathrm{s}^0\right)\mathcal{D}\frac{\gamma_\mathrm{m}}{2}}_{\text{dephasing}}
    + \underbrace{\sin\left(\tau\Delta_\mathrm{s}^0\right)\mathcal{C}\chi_\textrm{q--m}}_{\text{frequency shift}}
  \biggr],
\end{equation}
with the appropriate sign chosen such that $\eta \geq 0$. Based on the discussion in Section\;\ref{sec:theory-strong-dispersive-regime}, the frequency-shift term vanishes in the strong dispersive regime, but the dephasing term remains. In particular, the efficiency $\eta$ is always nonzero for $\tau > 0$ except at the nodes of the Ramsey fringes where $\cos(\tau\Delta_\mathrm{s}^0) = 0$~[Fig.\;4].

\subsection{Magnon sensing protocol and detection sensitivity}\label{sec:magnon-sensing-protocol}

The magnon sensing protocol uses the qubit as a quantum sensor, with the information about the magnon state conveyed through measurement of the qubit excited state probability $p_e$ via Ramsey interferometry of the qubit.
The signal $X$ in the protocol is the difference in $p_e$ between the cases of the magnon population $\overline{n}_\mathrm{m}$ being present and absent.
Specifically,
\begin{equation}\label{eq:X-full}
  X(\overline{n}_\mathrm{m}) = \left|p_e(\overline{n}_\mathrm{m}) - p_e(0)\right|.
\end{equation}
The efficiency $\eta$ shows its use here, as under the assumption of linearity in Eq.\;\eqref{eq:p-e-n-m-linear}, the signal $X$ is given by the very simple expression
\begin{equation}\label{eq:X-with-linear-response}
  X(\overline{n}_\mathrm{m}) = \lvert\eta\rvert\overline{n}_\mathrm{m}.
\end{equation}
The noise $\Xi_\mathrm{q}$ is obtained through Allan deviation analysis of repeated single-shot qubit measurements~[Fig.\;2(b)]. 
For the regime where Eq.\;\eqref{eq:p-e-n-m-linear} is valid, this can also be expressed with respect to the magnon population, $\Xi_\mathrm{m}$, with the adjustment
\begin{equation}
  \Xi_\mathrm{m} = \frac{\Xi_\mathrm{q}}{\eta}.
\end{equation} 
Due to the high readout fidelity, $\Xi_\mathrm{q}$ is dominated by shot noise, and can be expressed as~\cite{Degen2017}
\begin{equation}\label{eq:Xi-definition}
  \Xi_\mathrm{q}(T) = \sqrt{p_e(1-p_e)}\sqrt{\frac{1}{N(T)}},
\end{equation}
where $N$ is the number of shots and $T$ is the total time for all repetitions of the sequence.
This is equivalent to Eq.\;(3) in the main text, as $T=N\tau_\text{total}$.

The signal-to-noise ratio (SNR) is given by
\begin{equation}\label{eq:SNR-general-definition}
  \textrm{SNR}(T,\overline{n}_\mathrm{m},\ldots) = \frac{X(\overline{n}_\mathrm{m},\ldots)}{\Xi_\mathrm{q}(T,\overline{n}_\mathrm{m},\ldots)}. 
\end{equation}
By definition, the magnon detection sensitivity $\mathcal{S}$ is the smallest magnon population $\overline{n}_\mathrm{m}$ that can be detected with a unit SNR with a measurement time $T=\SI{1}{\second}$~\cite{Degen2017}.
As such, fixing the measurement time $T = \SI{1}{\second}$ and requiring $\textrm{SNR} = 1$ allow for finding the sensitivity, as in such a case
\begin{equation}\label{eq:nm-to-Snm}
  \overline{n}_\mathrm{m}\sqrt{T=\SI{1}{\second}}\Big\rvert_{\mathrm{SNR=1}} \equiv \mathcal{S}.
\end{equation}
An implicit relation for $\mathcal{S}$ can thus be found if the means to calculate $X$ and $\Xi_\mathrm{q}$ are known, as
\begin{equation}\label{eq:unit-snr-fraction}
  1 = \textrm{SNR}(T,\mathcal{S}/\sqrt{T},\ldots)\Big\rvert_{T=\SI{1}{\second}} = \frac{X(\mathcal{S}/\sqrt{T},\ldots)}{\Xi_\mathrm{q}(T,\mathcal{S}/\sqrt{T},\ldots)}\bigg\rvert_{T=\SI{1}{\second}}. 
\end{equation}

For small magnon populations $\nm\ll1$, the leading-order approximation of the noise $\Xi_\mathrm{q}$ is a constant, specifically
\begin{equation}
  \Xi_\mathrm{q} = \sqrt{p_e(\nm=0)(1-p_e(\nm=0))}\sqrt{\frac{\tau_\textrm{total}}{T}}.
\end{equation}
Under this assertion, the only dependence of the $\textrm{SNR}$ on $\nm$ is a linear dependence in $X$ according to Eq.\;\eqref{eq:X-with-linear-response}, which allows the use of Eq.\;\eqref{eq:unit-snr-fraction} to obtain a fully-analytical expression for the magnon detection sensitivity
\begin{equation}\label{eq:Snm-analytical}
  \Snm = \frac{\Xi_\mathrm{q}(T,\ldots)}{\lvert\eta\rvert}\sqrt{T}\Bigr|_{T=\SI{1}{\second}} = \frac{\sqrt{p_e(0)(1-p_e(0))}}{\lvert\eta\rvert}\sqrt{\tau_\textrm{total}}.
\end{equation}

\subsection{Sensitivity to microwave magnetic fields}

In addition to considering quantum sensing of a given magnon population $\nm$, the protocol can also be used to sense a microwave drive of constant amplitude. 
Such a drive would generate a magnetic field $B_\mathrm{d}$, which excites magnons in the Kittel mode.
The microwave magnetic field sensitivity $\Sbd$ can be defined analogously to $\Snm$, with $B_\mathrm{d}$ as the quantity of interest instead of $\nm$.

The relation between the magnon drive coefficient $\Omega_\mathrm{d}$ and the magnon population $\nm^g$ excited in the Kittel mode is given by
\begin{equation}\label{eq:n-m-Omega-d-relation}
  \nm^g = \frac{\Omega_\mathrm{d}^2}{(\gm/2)^2 + \Delta_\mathrm{d}^2},
\end{equation}
where $\gm$ is the magnon linewidth and $\Delta_\mathrm{d}$ is the detuning between the magnon drive and the magnon frequency.
The relation between the magnon drive coefficient $\Omega_\mathrm{d}$ and the amplitude of the microwave magnetic field $B_\mathrm{d}$ is \cite{Tabuchi2016}
\begin{equation}\label{eq:Omega-d-B-d-relation}
  \hbar\Omega_\mathrm{d} = g\mu_\mathrm{B}\frac{B_\mathrm{d}}{2}\sqrt{N_\textrm{spins}},
\end{equation}
where $g=2.002$ is the $g$-factor in YIG, $\mu_\mathrm{B}$ is the Bohr magneton, and $N_\text{spins}\approx\num{1.8e14}$ is the number of spins in the YIG sphere.
Assuming the magnon drive is resonant so $\Delta_\mathrm{d}=0$, this leads to
\begin{equation}\label{eq:B-d-n-m-relation}
  B_\mathrm{d} = \sqrt{\frac{\nm}{N_\textrm{spins}}}\left(\frac{\hbar\gm}{g \mu_\mathrm{B}}\right).
\end{equation}

After imposing the same conditions as are required for defining the sensitivity from the magnon population as in Eq.\;\eqref{eq:nm-to-Snm}, the sensitivity $\Sbd$ can be expressed in terms of $\Snm$ as
\begin{equation}
  \Sbd = B_\mathrm{d}\sqrt{T=\SI{1}{\second}}\Bigl|_{\textrm{SNR}=1}
  = \sqrt{\frac{\nm\left(\sqrt{T=\SI{1}{\second}}\right)^2}{N_\textrm{spins}}}\left(\frac{\hbar\gm}{g \mu_\mathrm{B}}\right)\Biggl|_{\textrm{SNR=1}}
  = \sqrt{\frac{\Snm\sqrt{T=\SI{1}{\second}}}{N_\textrm{spins}}}\left(\frac{\hbar\gm}{g \mu_\mathrm{B}}\right).
\end{equation}
Note that the additional factor of $\sqrt{T=\SI{1}{\second}}$ does not affect the numerical values when converting from $\Snm$ to $\Sbd$ when all other quantities are expressed in SI units, as is the case here.
For the results such as those quoted in the main text, $\Snm\approx\SI{1.5e-3}{\text{magnons}\per\sqrt{\Hz}}$ leads to $\Sbd\approx\SI{1}{\femto\tesla\per\sqrt{\Hz}}$ for a magnon linewidth \mbox{$\gm/2\pi\approx\SI{1.6}{\MHz}$}.

\section{Analytical and numerical modeling}

\subsection{Optimal magnon linewidth considering qubit dephasing due to magnons}\label{sec:optimal-magnon-linewidth}

\begin{figure*}[tb]
  \includegraphics{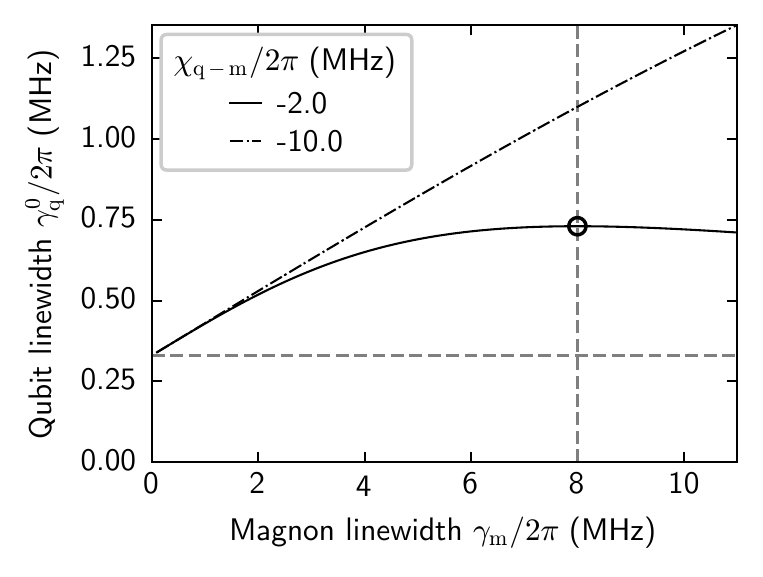}
  \caption{
  Qubit linewidth with the Kittel mode in the vacuum state,~$\gq^0/2\pi$, as a function of the magnon linewidth~$\gm/2\pi$ for different values of the qubit--magnon dispersive shift~$\chiqm/2\pi$, with the magnon drive taken to be resonant ($\Delta_\mathrm{d}=0$) and the magnon population fixed at~$\overline{n}_\mathrm{m}^g=0.1$~[Eq.\;\eqref{eq:optimal-magnon-linewidth}].
  The horizontal dashed line indicates the intrinsic qubit linewidth used in the simulations,~$\gq/2\pi=\SI{0.33}{\MHz}$.
  The vertical dashed line indicates a magnon linewidth of \mbox{$\gm/2\pi=\SI{8}{MHz}=4\lvert\chiqm\rvert/2\pi$} for the case of \mbox{$\chiqm/2\pi=\SI{-2}{\MHz}$}.
  The open circle marks the largest value of the qubit linewidth $\gq^0/2\pi$ for a dispersive shift of $\chiqm/2\pi=\SI{-2}{\MHz}$, which occurs at precisely $\gm/2\pi=\SI{8}{\MHz}$.
  }
  \label{fig:gamma-q-vs-gamma-m}
\end{figure*}

The qubit linewidth in Eq.\;\eqref{eq:qubit-lw-magnon-number} can be considered a proxy for the sensitivity, as the operating regime is such that the qubit response to the magnon population is dominated by dephasing.
In particular, it can serve to aid in examining optimization under a change of magnon linewidth.
In the case of a small magnon population $\nm^g\ll1$ so that the dominant qubit parameters correspond to the $n_\mathrm{m}=0$ Fock state, as well as a resonant magnon drive $\Delta_\mathrm{d} = 0$, the contribution to the qubit linewidth arising from the magnon population is given by
\begin{equation}\label{eq:optimal-magnon-linewidth}
  \begin{aligned}
    \gq^0(\nm^g) - \gq^0(0) &= \overline{n}_\mathrm{m}^g\mathcal{D}\gm \\
    &= \overline{n}_\mathrm{m}^g\gm\mathcal{A}(1+\mathcal{B})\\
    &= \overline{n}_\mathrm{m}^g\frac{4\chiqm^2\gm}{(\gm/2)^2+4\chiqm^2},
  \end{aligned}
\end{equation}
which reveals the dependence on the device parameters.
If the magnon population $\overline{n}_\mathrm{m}^g$ and dispersive shift $\chiqm$ are fixed, $\gq$ achieves a maximal value when $\gm=4\lvert\chiqm\rvert$~[Fig.\;\ref{fig:gamma-q-vs-gamma-m}].

Although this result is in principle exact, this is not a complete description of the condition for optimal sensitivity.
This is due to the fact that there is a small but nonzero shift in the frequency of the qubit described by Eq.\;\eqref{eq:qubit-freq-magnon-number} which also contributes to $p_e(\overline{n}_\mathrm{m}^g)$, so while the qubit linewidth is very closely tied to the magnon detection sensitivity, it does not fully determine the sensitivity by itself.

In contrast, when considering the sensitivity to a microwave magnetic field, the magnon drive amplitude must instead be fixed.
In such a case, combining the results of Eqs.\;\eqref{eq:B-d-n-m-relation} and \eqref{eq:optimal-magnon-linewidth} gives the dependence of the qubit linewidth on the magnon population as
\begin{equation}
  \begin{aligned}
    \gq^0(\nm^g) - \gq^0(0)
    &= \overline{n}_\mathrm{m}^g\frac{4\chiqm^2\gm}{(\gm/2)^2+4\chiqm^2} \\
    &= \frac{B_\mathrm{d}^2}{\gm^2}N_\textrm{spins}\left(\frac{g\mu_\mathrm{B}}{\hbar}\right)^2 \frac{4\chiqm^2\gm}{(\gm/2)^2+4\chiqm^2} \\
    &= B_\mathrm{d}^2 N_\textrm{spins}\left(\frac{g\mu_\mathrm{B}}{\hbar}\right)^2 
    \frac{4\chiqm^2}{\gm\left[(\gm/2)^2+4\chiqm^2\right]}.
  \end{aligned}
\end{equation}
For this expression, the maximum is achieved for $\gm\to0$ when all other parameters are kept fixed.

In conclusion, when maximizing the qubit linewidth, the optimal magnon linewidth is $\gm=4\lvert\chiqm\rvert$ for a fixed magnon population $\overline{n}_\mathrm{m}^g$ such as when considering the magnon detection sensitivity in terms of the magnon population.
This weak optimum can be understood by considering that a larger magnon linewidth, which itself leads to greater decoherence of the qubit, can also reduce the amount of information about the qubit state that is lost upon magnon decay. 
In particular, the greater the spectral overlap between the Kittel mode peaks corresponding to the qubit in the ground and excited states, the less the decaying magnons carry information about the qubit state, preserving the qubit coherence.
In contrast, there is no optimal magnon linewidth when considering the magnon detection sensitivity in terms of the amplitude of the microwave drive, as a smaller magnon linewidth leads to a larger magnon population that is excited by a drive of a given amplitude.
This change is enough to completely offset the effects discussed previously.

\subsection{Asymptotic limits of magnon detection sensitivity with respect to the magnon linewidth}\label{sec:asymptotic-limits}

Starting from the first-order approximation of the efficiency $\eta$ in Eq.\;\eqref{eq:eta-full}, the definitions of $\mathcal{C}$ and $\mathcal{D}$ can be used under the assumption of a resonant magnon drive $\Delta_\mathrm{d}=0$ to obtain
\begin{equation}
  \lvert\eta\rvert \propto \frac{\gm}{\gm^2+16\chiqm^2}
  \bigl[
    \underbrace{8\cos(\tau\Delta_\mathrm{s}^0)\chiqm^2}_\textrm{dephasing}
    + \underbrace{2\sin(\tau\Delta_\mathrm{s}^0)\gm}_\textrm{frequency shift}
  \bigr].
\end{equation}
For a resonant qubit drive $\Delta_\mathrm{s}^0=0$, the frequency-shift term vanishes, so the asymptotic scaling with respect to $\gm$ is determined purely by the prefactor.
An extremal value occurs at $\gm=4\lvert\chiqm\rvert$, at which point $\lvert\eta\rvert$ switches from increasing to decreasing for larger $\gm$.
This corresponds exactly to the discussion and conclusions in Section~\ref{sec:optimal-magnon-linewidth}, which considered only the qubit linewidth as a proxy for the efficiency and sensitivity.
  
Examining the change in scaling for very small and very large values of $\gm$,
\begin{align}
  \lim_{(\gm/\chiqm)\to0}\bigl|\eta\bigr| &\propto \frac{\gm}{16\chiqm^2} + \mathcal{O}(\gm^3) \sim \gm, \\
  \lim_{(\gm/\chiqm)\to\infty} \bigl|\eta\bigr| &\propto  \frac{1}{\gm} + \mathcal{O}\left(\left(\frac{1}{\gm}\right)^3\right) \sim \frac{1}{\gm}.
\end{align}
This implies that the asypmtotic limits for small and large $\gm$ have gradients of different signs, which necessitates the existence of an extremal value for $\eta$.
Extending this to the magnon-number sensitivity $\Snm$ using Eq.\;\eqref{eq:Snm-analytical}, the sensitivity follows $\Snm\propto\lvert1/\eta\rvert$ so the existence of an optimal magnon linewidth is almost directly reflected in $\Snm$ as well, as
\begin{align}
  \lim_{(\gm/\chiqm)\to0}\bigl[\Snm\bigr] &\propto \lim_{(\gm/\chiqm)\to0}\left|\frac{1}{\eta}\right| \sim \frac{1}{\gm}, \\
  \lim_{(\gm/\chiqm)\to\infty}\bigl[\Snm\bigr] &\propto \lim_{(\gm/\chiqm)\to\infty}\left|\frac{1}{\eta}\right| \sim \gm.
\end{align}

For the sensitivity to microwave magnetic field $\Sbd$, Eq.\;\eqref{eq:B-d-n-m-relation} shows that
\begin{equation}
  \Sbd \propto \sqrt{\Snm}\gm 
  \propto \sqrt{\frac{\gm^2 + 16\chiqm^2}{8\cos(\tau\Delta_\mathrm{s}^0)\chiqm^2}}\sqrt{\gm},
\end{equation}
which clearly has no optimal value with respect to $\gm$ and will always increase.
Explicitly,
\begin{align}
  \lim_{(\gm/\chiqm)\to0}\bigl[\Sbd\bigr] &\propto \lim_{(\gm/\chiqm)\to0}\bigl[\sqrt{\Snm}\gm\bigr] \sim \frac{\gm}{\sqrt{\gm}} = \sqrt{\gm}, \\
  \lim_{(\gm/\chiqm)\to\infty}\bigl[\Sbd\bigr] &\propto \lim_{(\gm/\chiqm)\to\infty}\bigl[\sqrt{\Snm}\gm\bigr] \sim \sqrt{\gm}\gm = \gm^{3/2}.
\end{align}
This can be understood by considering the fact that increasing the magnon linewidth decreases the effectiveness of a fixed microwave magnetic field, resulting in a smaller magnon population.
This outweighs the effects that lead to a finite optimal magnon linewidth for the sensitivity with respect to a fixed magnon number $\Snm$, resulting in an optimal magnon linewidth of zero for $\Sbd$.


\subsection{Governing equations for numerical simulations}

Numerical simulations are based on the doubly-rotating frame effective Hamiltonian~\cite{Lachance-Quirion2020}
\begin{equation}\label{eq:sim-hamiltonian-full}
  \begin{aligned}
    \hat{\mathcal{H}}_\mathrm{eff}(t)/\hbar =& 
    \left(\Delta_\mathrm{s} - \frac{\alpha}{2}\right)\hat{b}^\dag\hat{b} + \frac{\alpha}{2}\left(\hat{b}^\dag\hat{b}\right)^2
    + \Delta_\mathrm{d}\hat{c}^\dag\hat{c}
    + 2\chi_\mathrm{q-m}\hat{b}^\dag\hat{b}\hat{c}^\dag\hat{c} \\
    &+ \Omega_\mathrm{s}(t)\left(\hat{b}^\dag + \hat{b}\right)
    + \Omega_\mathrm{d}(t)\Bigl[\hat{c}^\dag + \hat{c}
    + \Lambda_{ef} \bigl(\ket{e}\bra{f} + \ket{f}\bra{e}\bigr)\Bigr],
  \end{aligned}
\end{equation}
with $\Omega_\mathrm{s}(t)$ and $\Omega_\mathrm{d}(t)$ the time-dependent excitation strengths for the qubit and Kittel mode, respectively.
The coefficient $\Lambda_{ef}$ represents the magnon drive exciting the qubit $\ket{e}\leftrightarrow\ket{f}$ transition, and its inclusion is motivated by the fact that the detuning of $\SI{18}{\MHz}$ between the Kittel mode and the qubit $\ket{e}\leftrightarrow\ket{f}$ transition is comparable to the qubit--magnon coupling strength $g_\textrm{q--m}/2\pi=\SI{7.07}{\MHz}$. 
The predicted value of $\Lambda_{ef}$ can be calculated by considering a drive of power $P_\mathrm{d}$ applied at frequency $\omega_\mathrm{d}$, and calculating the resulting excitation strengths $\Omega_\mathrm{d}$ and $\Omega_{ef}=\Omega_\mathrm{d}\Lambda_{ef}$ from input-output theory.
The magnon excitation strength is given by \cite{Lachance-Quirion2017}
\begin{equation}\label{eq:magnon-excitation-strength}
  \Omega_\mathrm{d} = \sqrt{\frac{P}{\hbar\omega_\mathrm{d}}}
  \sum_p \sqrt{\kappa_\cavp^\textrm{in}}\left[
    \frac{g_{\textrm{m--}\cavp}}{\Delta_{\textrm{m--}\cavp}} 
    + \frac{g_\textrm{q--m}g_{\textrm{q--}\cavp}}{\Delta_\textrm{q--m}\sqrt{\Delta_{\textrm{m--}\cavp}^2 + \kappa_\cavp^2}}
    \right],
\end{equation}
where $\Delta_{\textrm{m--}\cavp}\equiv\omega_\mathrm{m}-\omega_\cavp$ is the detuning between the Kittel mode and $\mathrm{TE}_{10p}$ cavity mode, and $\kappa_\cavp^\textrm{in}$ is the coupling rate of the input port of the cavity to the $\mathrm{TE}_{10p}$ cavity mode.
The qubit $\ket{e}\leftrightarrow\ket{f}$ excitation strength is given by
\begin{equation}\label{eq:qubit-ef-excitation-strength}
  \Omega_{ef} = \sqrt{\frac{P}{\hbar\omega_\mathrm{d}}}
  \sum_p \sqrt{\kappa_\cavp^\textrm{in}}\left[
    \frac{\sqrt{2}g_{\textrm{q--}\cavp}}{\omega_\cavp-\omega_{ef}}
    \right],
\end{equation}
where it is assumed that $g_{ef\textrm{--}\cavp} = \sqrt{2}g_{\textrm{q--}\cavp}$.
The ratio of the excitation strengths is then \mbox{$\Lambda_{ef} = \Omega_{ef}/\Omega_\mathrm{d}$}.

\subsection{Simulated sequences}\label{sec:simulated-sequences}

The numerical simulations are carried out using \textsc{QuTiP} \cite{Johansson2013}, using device parameters obtained from characterization measurements~[Table~\ref{tab:sim-param-values}].
The processes, rates, and operators used in the Lindblad master equation are defined in Table~\ref{tab:sim-collapse-ops}.
In general, the simulations attempt to faithfully reproduce the measurements carried out in the experiment~[Section~\ref{sec:experiments}].
The qubit thermal population is set via the expectation value $\nm^\text{th}=0.038$, so that for a three-level transmon, the thermal excited state probability $p_e^\text{th} = 0.035$ matches that determined from the experiments.
The high-power readout of the qubit is not simulated due to the excessive computational cost of simulating $>10^4$ photons in the cavity.
Instead, the instantaneous expectation values of the probabilities of finding the qubit in the state~$i$ at the readout time~$t_r$, $p_i(t=t_r)$ for $i=g,e,f$, are used, and are adjusted for readout errors as in Ref.\;\cite{Lachance-Quirion2020}.
The Kittel mode linewidth $\gm/2\pi=\SI{1.567}{\MHz}$ is taken from the fitting of the qubit spectrum with an applied magnon drive~[Fig.\;\ref{fig:magnon-number-splitting-exp}(f)], as this matches the experimental conditions used to characterize the magnon detection sensitivity, in particular the applied current and therefore the Kittel mode frequency.

\begin{table*}[tb]
  \caption{
    Parameters and values used in numerical simulations and analytical models. For values in the ``Numerics'' column, uncertainties are not used in the simulations, but are quoted for quantities which originate from analysis of multiple measurements to give an indication of their variation over the whole dataset.}
  \begin{ruledtabular}
  \begin{tabular*}{\textwidth}{ll@{\extracolsep{\fill}}d{9}d{3}c}
    Parameter & Unit & \multicolumn{1}{c}{Numerics} & \multicolumn{1}{c}{Analytics} & Figure \\
    \hline
    Dressed qubit anharmonicity $\alpha_0/2\pi$ &$\si{MHz}$ & -122.61 & - & $-$ \\
    Magnon linewidth $\gamma_\mathrm{m}/2\pi$ &$\si{MHz}$ & 1.567\,(41) & - & \ref{fig:magnon-number-splitting-exp} \\
    Qubit linewidth $\gamma_\mathrm{q}^0/2\pi$ &$\si{MHz}$ & 0.330\,(4) & 0.354 & \ref{fig:qubit-char}(b) \\
    Qubit--magnon dispersive shift $\chi_\textrm{q--m}/2\pi$ &$\si{MHz}$ & -1.762\,(22) & - & \ref{fig:magnon-number-splitting-exp} \\
    \hline
    Qubit thermal population $\overline{n}_\mathrm{q}^\mathrm{th}$ && 0.038 & - & $-$ \\
    Magnon thermal population $\overline{n}_\mathrm{m}^\mathrm{th}$ && 0.0 & - & $-$ \\
    \hline
    Average qubit drive detuning $\Delta_\mathrm{s}^0/2\pi$ &$\si{kHz}$ & 16\,(32) & 0 & \ref{fig:qubit-char}(b) \\
    Average magnon drive detuning $\Delta_\mathrm{d}/2\pi$ &$\si{kHz}$ & -42\,(8) & 0 & \ref{fig:magnon-spectroscopy} \\
    Qubit drive frequency $\omega_\mathrm{s}/2\pi$ &$\si{\GHz}$ & 7.914\,446 & - & $-$ \\
    Magnon drive frequency $\omega_\mathrm{d}/2\pi$ &$\si{\GHz}$ & 7.781\,015 & - & $-$ \\
    \hline
    Qubit relaxation time $T_1$ &$\si{\mu\second}$ & 0.801\,(14) & - & \ref{fig:qubit-char}(a) \\
    Qubit coherence time $T_2^\ast$ &$\si{\mu\second}$ & 0.965\,(10) & 0.9 & \ref{fig:qubit-char}(b) \\
    \hline
    Minimal excited state probability $p_e^{\ket{g}}$ && 0.0860\,(28) & 0 & \ref{fig:qubit-readout}(a) \\
    Maximal excited state probability $p_e^{\ket{e}}$ && 0.8380\,(73) & 1 & \ref{fig:qubit-readout}(b) \\
    \hline
    Calibration target $\overline{n}_\mathrm{m}$ && 0.615\,(12) & - & \ref{fig:magnon-number-splitting-exp} \\
    Excitation target $\overline{n}_\mathrm{m}$ && \multicolumn{2}{c}{$0$ to $0.05$, 6 values} & $-$ \\
  \end{tabular*}
  \end{ruledtabular}
  \label{tab:sim-param-values}
\end{table*}

\begin{table*}[tb]
  \caption{Lindblad master equation processes and rates considered in numerical simulations.}
  \begin{ruledtabular}
  \begin{tabular*}{\textwidth}{l@{\extracolsep{\fill}}ll}
    Process & Rate $\gamma_k$ & Operator $\hat{L}_k$ \\
    \hline
    Qubit relaxation & $\gamma_{1}\left(1+\overline{n}_\mathrm{q}^\mathrm{th}\right)$ & $\hat{b}$ \\
    Qubit thermal excitation & $\gamma_{1}\overline{n}_\mathrm{q}^\mathrm{th}$ & $\hat{b}^\dag$ \\
    Qubit pure dephasing & $2\gamma_{\varphi}$ & $\hat{b}^\dag\hat{b}$ \\
    \hline
    Magnon relaxation & $\gamma_\mathrm{m}\left(1+\overline{n}_\mathrm{m}^\mathrm{th}\right)$ & $\hat{c}$ \\
    Magnon thermal excitation & $\gamma_\mathrm{m}\overline{n}_\mathrm{m}^\mathrm{th}$ & $\hat{c}^\dag$ \\
  \end{tabular*}
  \end{ruledtabular}
  \label{tab:sim-collapse-ops}
\end{table*}

To calibrate the qubit control pulse amplitude corresponding to a $\pi$ rotation, a Gaussian pulse with amplitude~$\Omega_\mathrm{s}$~[Eq.\;\ref{eq:qubit-pulse-envelope}] is applied to the qubit, for various values of $\Omega_\mathrm{s}$.
The $\pi$-pulse amplitude $\Omega_\mathrm{s}^\pi$ is obtained by determining the value of $\Omega_\mathrm{s}$ which maximizes the  probabilty $p_e(t_\mathrm{r})$ of finding the qubit in the excited state at the readout time~$t_\mathrm{r}$. 
As readout errors do not affect this calibration, the readout time $t_\mathrm{r}$ is taken as the start of the readout pulse.
The simulation is repeated with a pulse of amplitude $\Omega_\mathrm{s}^\pi$ for various readout times $t_\mathrm{r}$. 
Estimation of the appropriate instantaneous readout time is carried out as described in Ref.\;\cite{Lachance-Quirion2020}, with the best estimate $\Delta t_\mathrm{r}=\SI{32}{\nano\second}$ after the start of the readout pulse. 

Ramsey interferometry of the qubit is carried out by two pulses of amplitude $\Omega_\mathrm{s}^\pi/2$, reproducing the timing from the experiment.
The intrinsic qubit linewidth $\gamma_\mathrm{q}/2\pi=\SI{0.33}{\MHz}$ is used such that the decoherence time $T_2^\ast$ obtained from fitting the simulation data matches the value of $T_2^\ast=\SI{0.89}{\mu\second}$ obtained from fitting the experimental data.

\begin{figure*}[tb]
  \includegraphics[]{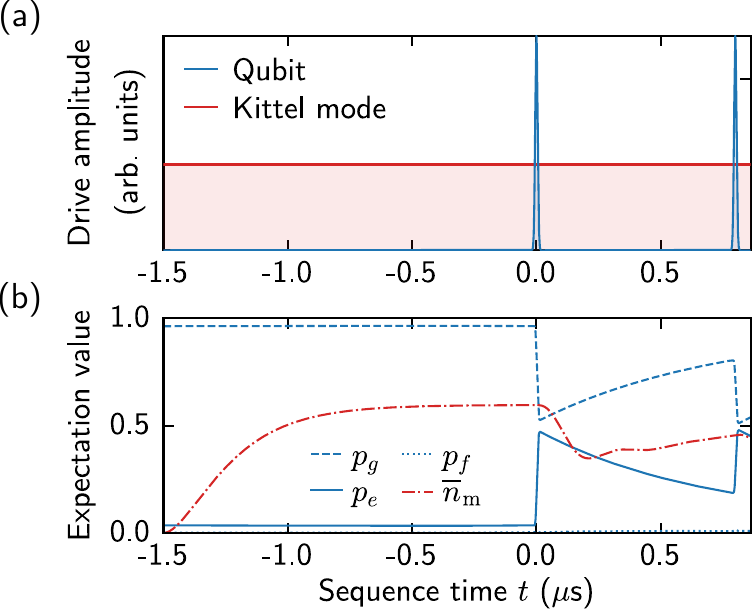}
  \caption{
  Simulation time traces for magnon drive calibration sequence with $\tau=\SI{800}{\nano\second}$.
  (a) Drives applied to the transmon qubit~(blue) and Kittel mode~(red).
  (b) Probabilities $p_i$ for the three transmon states~$\ket{g}, \ket{e}, \ket{f}$, and the average magnon number $\overline{n}_\mathrm{m}$, throughout the sequence simulation.
  Time $t=0$ is defined as the center of the first qubit pulse. The simulation starts at $t=\SI{-1.5}{\mu\second}$ to allow the magnon population to reach a steady state and so mimic the continuous drive applied in the experiment.
  The readout occurs at the upper limit of the time axis, here at $t = \SI{0.862}{\mu\second}$.
  The qubit $\ket{f}$ state population is barely visible at this scale.
  }
  \label{fig:simulation-traces}
\end{figure*}

To calibrate the magnon population excited for a given magnon drive amplitude, the experimental procedure described in Sections~\ref{sec:strong-dispersive-regime} and \ref{sec:calibration-magnon-population} is reproduced in the simulations~[Fig.\;\ref{fig:simulation-traces}].
In the case of numerical simulations, the magnon drive amplitude is given by $\Omega_\mathrm{d}$, which is varied in order for the fitting of the resulting qubit spectrum to yield the same magnon population as was found in the experiment, $\overline{n}_\mathrm{m}=0.615$. 
When characterizing the magnon detection sensitivity, the magnon population from fitting is taken directly from the fit of the model in Section~\ref{sec:strong-dispersive-regime} to the qubit excited state probability $p_e$ at the measurement time, and thus exactly matches the procedure used in the experiments. 
Importantly, the magnon population experiences oscillatory variations during the sensing time, caused by the dispersive shift of the Kittel mode away from the magnon drive frequency due to the population of the qubit excited state during the sequence~[Fig.\;\ref{fig:simulation-traces}(b)].
As the analytical models, such as those whose results are shown in Fig.\;1(b), use $\nm^g$ as the definition of the magnon population, this dynamical detuning leads to a discrepancy between the results of the analytical models and those of the experiments and numerical simulations.

\begin{figure*}[tb]
  \includegraphics[]{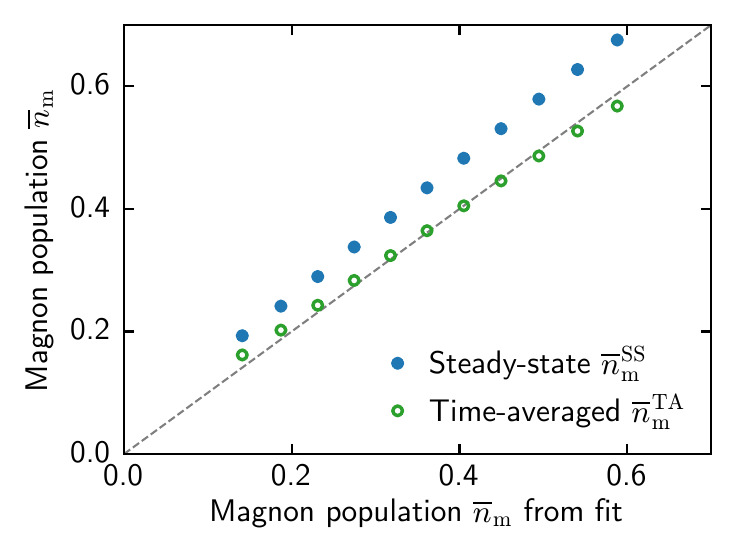}
  \caption{
  Comparison of different values of the magnon population $\nm$ from simulations of the magnon sensing sequence, showing the steady-state population $\nm^\text{SS}$ (blue dots) and the time-averaged population $\nm^\text{TA}$ (green circles).
  Both are shown as a function of the value of $\overline{n}_\mathrm{m}$ obtained from fitting the qubit spectrum in the same way as is done for the experimental data.
  The gray dashed line is a unit slope.
  }
  \label{fig:simulated-n-m-comparison}
\end{figure*}

To highlight this difference, two alternative means of determining the magnon population in the numerical simulations are considered.
The steady-state magnon population, $\overline{n}_\mathrm{m}^\text{SS}$, is defined as the magnon population generated by the magnon drive after the ramp-up time but before any pulses are applied to the qubit. 
Specifically, considering the magnon population, $\overline{n}_\mathrm{m}(t)$, at a given time $t$ in the simulation, 
\begin{equation}
  \overline{n}_\mathrm{m}^\text{SS} = \overline{n}_\mathrm{m}(t=\SI{-18}{\nano\second}),
\end{equation}
as the first qubit pulse begins at $t=\SI{-18}{\nano\second}$~[Fig.\;\ref{fig:simulation-traces}].
The time-averaged magnon population, $\overline{n}_\mathrm{m}^\text{TA}$, is defined as the average of the magnon population $\overline{n}_\mathrm{m}(t)$ for all times $t$ within the sensing time $\tau$ evaluated during the simulation for a given set of parameters. In particular, as the qubit spectrum with which the magnon population is calibrated is obtained from a set of results for different $\tau$ values, 
\begin{equation}
  \overline{n}_\mathrm{m}^\text{TA} 
  = \frac{1}{\left(\sum_{\ell=0}^{\tau_\text{max}/\Delta\tau}\ell\Delta\tau\right)} 
    \left[\sum_{\ell=0}^{(\tau_\text{max}/\Delta\tau)}\sum_{m=0}^{(\ell\Delta\tau/\Delta t)} \overline{n}_\mathrm{m}(m\Delta t)\right],
\end{equation}
where $\tau_\text{max} = \SI{4}{\mu\second}$ is the longest sensing time used, $\Delta\tau = \SI{40}{\nano\second}$ is the step size used with variation of sensing time, and $\Delta t=\SI{0.1}{\nano\second}$ is the step size used for time evolution of the system in the simulation.  
Figure~\ref{fig:simulated-n-m-comparison} compares $\nm^\text{SS}$ and $\nm^\text{TA}$ values against $\nm$ obtained from the fit of the qubit spectrum for different amplitudes of the magnon drive.
As the population $\nm$ from the fitting of the spectrum is much more closely aligned with the time-averaged population $\nm^\text{TA}$, it is evident that the sensing protocol detects the average magnon population during the sequence moreso than the population before the sequence begins.
Nevertheless, this discrepancy is at most $\sim10\%$ for larger values of $\nm$, and is suppressed as the value approaches $\nm=0$.

\begin{figure*}[tb]
  \includegraphics[]{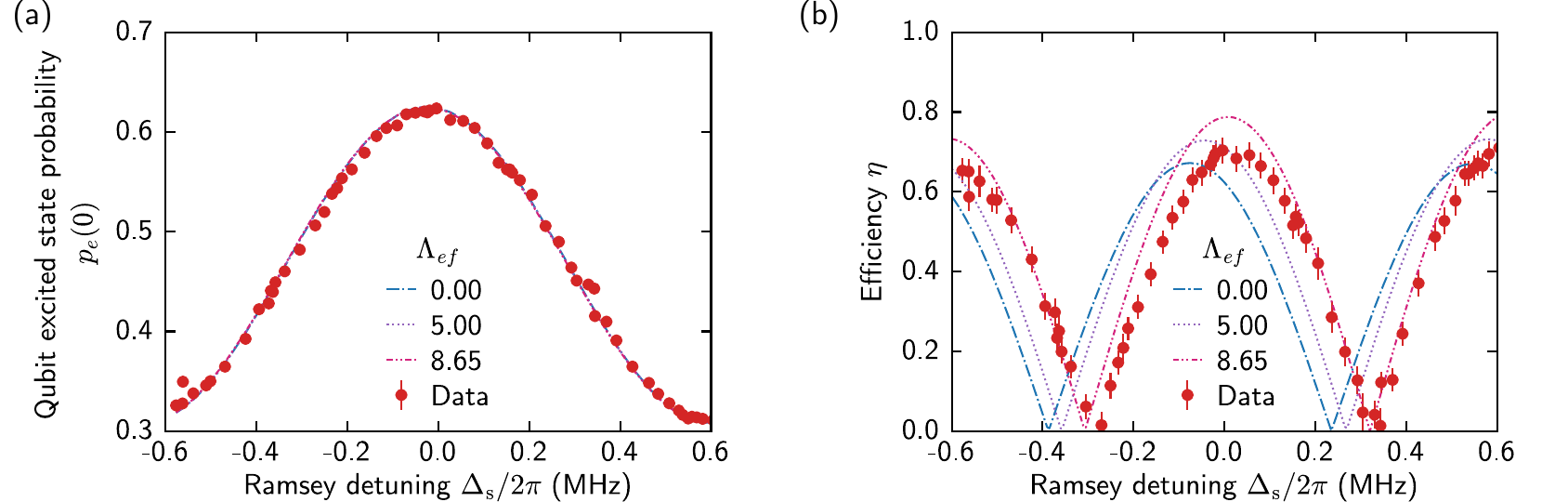}
  \caption{
  Comparison of numerical simulations of (a) the qubit excited state probability, $p_e(0)$, after the sensing sequence with no magnons present, and (b) the efficiency $\eta$ as a function of the Ramsey detuning $\Delta_\mathrm{s}/2\pi$ for several values of $\Lambda_{ef}$ (lines), as well as data from dataset~2 of the experiments (red dots).
  The coefficient $\Lambda_{ef}$ quantifies the ratio of the magnon pump power that unintentionally drives the qubit $\ket{e}\leftrightarrow\ket{f}$ transition, shifting the frequency dependence of the efficiency.
  The optimal value $\Lambda_{ef}=8.65$ minimizes the sum of squared differences between the efficiency~$\eta$ for the simulation and data.
  }
  \label{fig:xdrive-comparison}
\end{figure*}

The efficiency $\eta$ is simulated by reproducing the experimental procedure with respect to the range of target values of the magnon population $\nm$, and fitting to Eq.\;(1).
The noise $\Xi_\mathrm{q}$ is calculated using the shot noise model~[Eq.\;(3)].
The magnon-drive-induced $\ket{e}\leftrightarrow\ket{f}$ excitation coefficient $\Lambda_{ef}$ is included in the sequences to calibrate the magnon drive amplitude, as well as the sequences to calibrate the magnon detection sensitivity.
It is the only free parameter used in the model.
The optimized value $\Lambda_{ef} = 8.650$ is obtained by minimizing the sum of squared differences between the efficiency~$\eta$ obtained from the experiments in dataset~2 and the simulations with the same parameters~[Fig.\;\ref{fig:xdrive-comparison}].
This shows that the sensitivity is not significantly affected, and the effect of this drive is primarily equivalent to a frequency shift.
The value of $\Lambda_{ef}$ can also be calculated analytically by considering the ratios of the excitation strengths in Eqs.\;\eqref{eq:magnon-excitation-strength} and \eqref{eq:qubit-ef-excitation-strength}.
Using values in Tables~\ref{tab:params-cavity} and \ref{tab:params-other}, considering only the lowest two cavity modes results in $\Lambda_{ef}=3.48$, while considering three yields $\Lambda_{ef}=4.41$, so the number of cavity modes included has a significant effect on the calculated value.
As higher-order cavity modes that are not included in the calculations are thus expected to have a slight influence on the final result, the value obtained from optimization with respect to the experimental data is seen to be a reasonable estimate.

\subsection{Comparing analytical and numerical models}

\begin{figure*}[tb]
  \includegraphics[]{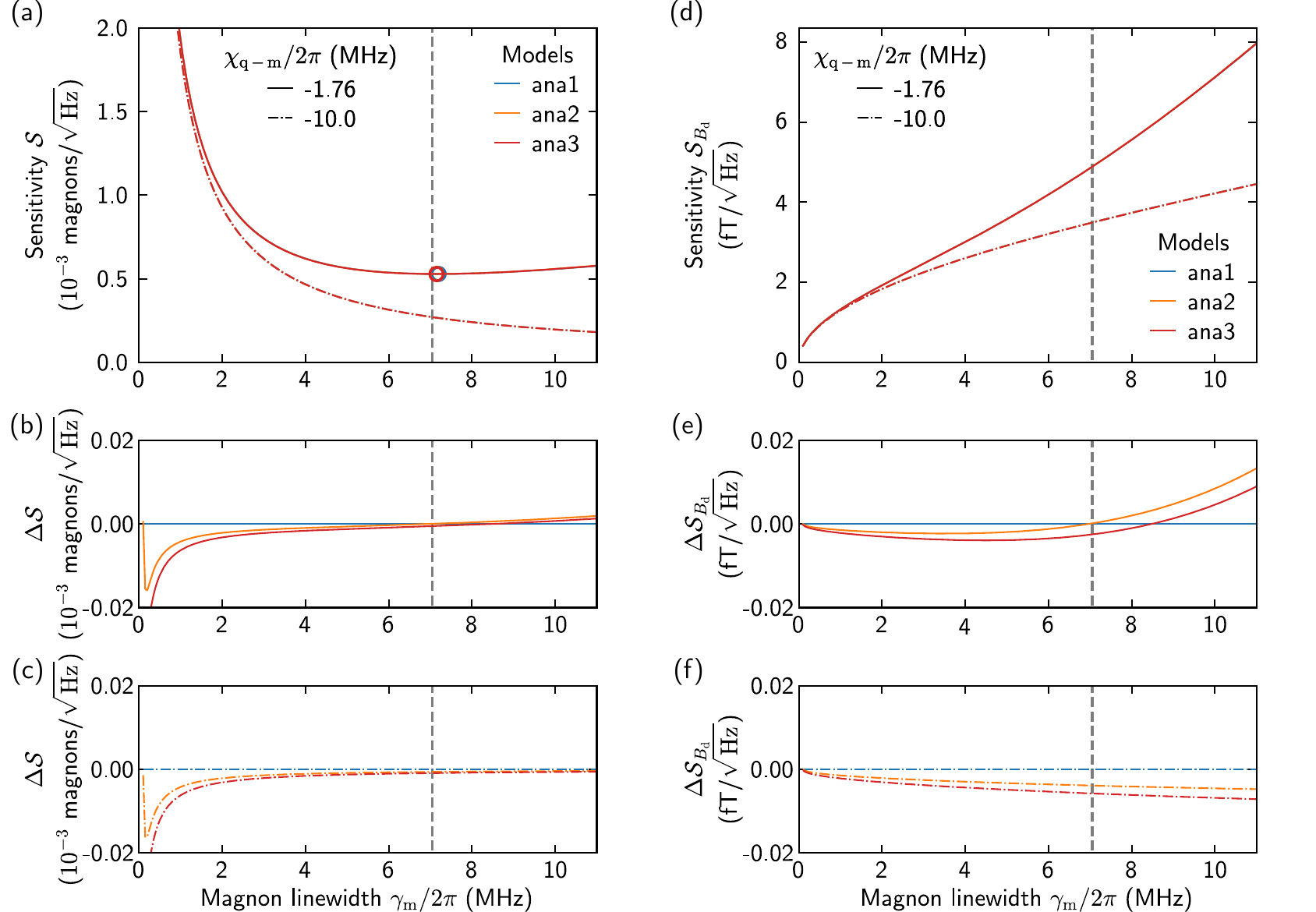}
  \caption{
  (a) Magnon detection sensitivity~$\Snm$ for different analytical models, as a function of magnon linewidth $\gm/2\pi$. 
  Also shown are differences~$\Delta\Snm$ relative to the model \texttt{ana1} for the dispersive shift (b) $\chiqm/2\pi=\SI{-1.76}{\MHz}$, which corresponds to the actual dispersive shift in the experiment, and (c) $\chiqm/2\pi=\SI{-10}{\MHz}$.
  Circles in~(a) indicate the optimal sensitivity~$\mathcal{S}$ for a given value of $\chiqm$. 
  The optimal values for $\chiqm/2\pi=\SI{-10}{\MHz}$ are not visible over the range in the figure.
  The gray vertical dashed line corresponds to $\gm=4\lvert\chiqm\rvert$ for the dispersive shift from the experiment.
  (d-f)~Corresponding comparisons for the microwave magnetic field sensitivity~$\Sbd$.
  In this case, the optimal sensitivity values are not indicated as the sensitivity always improves for $\gm\to0$.
  }
  \label{fig:analytical-comparison}
\end{figure*}

\begin{figure*}[tb]
  \includegraphics[]{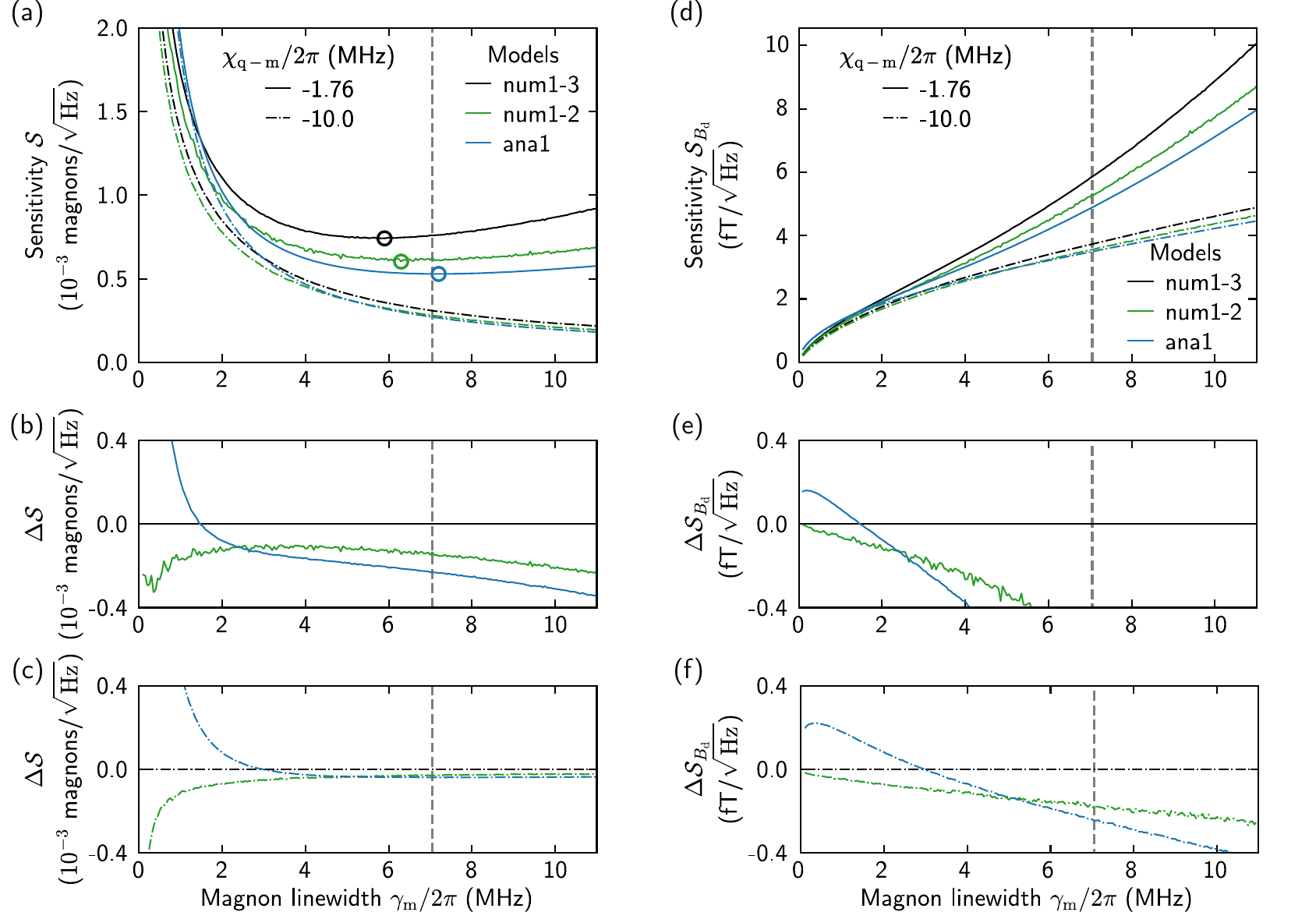}
  \caption{
  (a) Magnon detection sensitivity~$\Snm$ for numerical models with two and three qubit states, \texttt{num1-2} and \texttt{num1-3} respectively, and the \texttt{ana1} analytical model, as a function of magnon linewidth $\gm/2\pi$. 
  Also shown are differences~$\Delta\Snm$ relative to the model \texttt{num1-3} for the dispersive shift (b)~$\chiqm/2\pi=\SI{-1.76}{\MHz}$, which corresponds to the actual dispersive shift in the experiment, and (c)~$\chiqm/2\pi=\SI{-10}{\MHz}$.
  Circles represent the optimal sensitivity~$\mathcal{S}$ for a given value of $\chiqm$. 
  The gray vertical dashed line corresponds to $\gm=4\lvert\chiqm\rvert$ for the dispersive shift from the experiment.
  The optimal values for $\chiqm/2\pi=\SI{-10}{\MHz}$ are not visible over the range in the figure.
  (d-f)~Corresponding comparisons for the microwave magnetic field sensitivity~$\Sbd$.
  In this case, the optimal sensitivity values are not indicated as the sensitivity always improves for $\gm\to0$.
  }
  \label{fig:numerical-comparison}
\end{figure*}

To highlight the impact of different approximations such as those discussed in Section~\ref{sec:magnon-sensing-protocol}, the sensitivities $\Snm$ and $\Sbd$ are calculated using various analytical and numerical models that represent different levels of approximation.
The parameter values used are those listed in Table~\ref{tab:sim-param-values}, with a qubit drive detuning of $\Delta_\mathrm{s}=0$ and a sensing time $\tau=\SI{0.8}{\mu\second}$.
Additionally, errors associated only with the qubit readout are ignored, and the noise $\Xi_\mathrm{q}$ is taken to consist solely of shot noise.

A robust technique for finding the sensitivity involves considering the function 
\begin{equation}
  f(\nm,\ldots) \equiv \textrm{SNR}(T,\nm,\ldots)\Big\rvert_{T=\SI{1}{\second}} - 1 = \frac{X(\nm,\ldots)}{\Xi_\mathrm{q}(T,\nm,\ldots)}\bigg\rvert_{T=\SI{1}{\second}} - 1.
\end{equation}
With this definition, the sensitivity $\Snm$ can be calculated by finding a root of $f$ with respect to $\nm$, as by construction \mbox{$f(\Snm/\sqrt{T},\ldots)\rvert_{T=\SI{1}{\second}}=0$}~[Eq.\;\eqref{eq:nm-to-Snm}].
This allows the definition of several models as follows:
\begin{itemize}
  \item The analytical model \texttt{ana1} considers $p_e(\nm)$ and $p_e(0)$ calculated using the most complete analytical expressions for the qubit probability after a Ramsey sequence in the presence of a magnon population~[Eq.\;\eqref{eq:p-e-magnon-number}]. The signal $X$ and noise $\Xi$ are calculated according to their most fundamental analytical definitions in Eqs.\;\eqref{eq:X-full} and \eqref{eq:Xi-definition}, respectively, and these definitions are used to construct $f$. This is the least approximate analytical model examined here.
  \item For the more approximate analytical model \texttt{ana2}, the assumption is made that the signal $X$ is linear with respect to $\nm$, allowing the efficiency $\eta$ to be defined through Eq.\;\eqref{eq:X-with-linear-response} and calculated according to Eq.\;\eqref{eq:eta-full}. The noise $\Xi_\mathrm{q}$ remains dependent on $p_e(\nm)$, which is calculated via $\eta$. The function $f$ is constructed as for \texttt{ana1}. This model is the one that corresponds most closely to the assertions made when carrying out analysis of the experimental results.
  \item The simplest analytical model \texttt{ana3} further assumes that $\Xi_\mathrm{q}$ is independent of $\nm$ and can be calculated considering only $p_e(0)$. Uniquely, this allows the sensitivity $\Snm$ to be calculated directly using Eq.\;\eqref{eq:Snm-analytical} instead of via finding a root of $f$.
\end{itemize}
Figure~\ref{fig:analytical-comparison} shows the three analytical models compared as a function of the magnon linewidth $\gm$, for two values of the qubit--magnon dispersive shift~$\chiqm$.
It is clear that, once $\gm$ is sufficiently large, the three models yield barely-distinguishable results.
Importantly, this validates the assertion of linear qubit response used in the analysis of the experimental data.
Additionally, by considering the variation of the sensitivities with respect to the magnon linewidth $\gm$, the conclusions of Sections~\ref{sec:optimal-magnon-linewidth} and \ref{sec:asymptotic-limits} are verified while also revealing the behaviour for intermediate values of $\gm$.
Specifically, the optimal magnon linewidth is seen to be almost exactly $\gm = 4\lvert\chiqm\rvert$ for all three models.

It is difficult to directly compare the analytical models to the experimental data and full numerical simulations as it is not clear how to incorporate the additional driving of the qubit $\ket{e}\leftrightarrow\ket{f}$ transition.
Nevertheless, it is useful to compare the analytical models to numerical simulations considered without this additional drive, as this yields insights into what factors contribute to the discrepancies between the theory and experimental results in general.
It is important to note that the analytical models do account for the distinction between $\nm^g$ and $\nm^e$, which is not included in the numerical simulations or experimental analysis. 
On the other hand, the numerical simulations can account for the second excited state $\ket{f}$ of the transmon qubit, as well as naturally incorporating time-dependent effects such as a dynamical detuning of the Kittel mode and the qubit during the sensing sequence.

A simplified numerical simulation is used for this purpose, as the fitting of the magnon-number peaks of the qubit spectrum, according to the magnon population calibration procedure presented in Section~\ref{sec:strong-dispersive-regime}, gives poor results for large values of $\gm$ and small magnon populations $\nm$.
The steady-state magnon population $\nm$ is instead found using \textsc{QuTiP}'s steady-state solver, considering the base Hamiltonian involving the time-independent terms in Eq.\;\eqref{eq:sim-hamiltonian-full} as well as the magnon driving term with $\Omega_\mathrm{d}(t) = \Omega_\mathrm{d}$.
This approach is valid as there is a monotonic bijective relationship between $\Omega_\mathrm{d}$ and $\nm$ for sufficiently small drive strengths when all other parameters are fixed.
Here, the same procedure is used as in the \texttt{ana1} and \texttt{ana2} models to calculate the sensitivity~$\Snm$ from the qubit excited state probability~$p_e(\nm)$ via finding a root of the function $f$.
In this case, however, the value of $p_e(\nm)$ is calculated by numerically simulating the sequence according to the Hamiltonian in Eq.\;\eqref{eq:sim-hamiltonian-full} with the qubit $\ket{e}\leftrightarrow\ket{f}$ drive coefficient fixed at $\Lambda_{ef}=0$.

Figure~\ref{fig:numerical-comparison} shows the simplified numerical simulations carried out considering a transmon (ideal qubit) with a Hilbert space size $N_\mathrm{q}=3$ ($N_\mathrm{q}=2$) labeled \texttt{num1-3} (\texttt{num1-2}), additionally compared to the analytical model \texttt{ana1}.
The numerical models for both the transmon and ideal qubit converge for small values of $\gm$, however, large values of $\gm$ result in a better agreement between the results for the ideal qubit and analytical model, while the three-level transmon result is different.
This discrepancy is approximately on the order of the discrepancy between the experiment and full numerical simulations~[Figs.\;3c and 4b], therefore suggesting that the inclusion or exclusion of higher-order excited states of the qubit may play a role in determining the accuracy of the theory.
Additionally, this shows that the experimental results may be improved by using techniques to minimize the population of the $\ket{f}$ state, such as using longer qubit excitation pulses for better frequency selection, or pulse shaping such as DRAG~\cite{Motzoi2009}.
This also supports the motivation to include the $\ket{e}\leftrightarrow\ket{f}$ driving term in the full numerical simulation, as the existence of this transition and a nonzero population of the $\ket{f}$ state have a small but significant effect on the final result.
Overall, the numerical and analytical results show agreement in terms of the essential features of the behaviour as a function of $\gm$, and in particular reinforce the conclusions of Sections~\ref{sec:optimal-magnon-linewidth} and \ref{sec:asymptotic-limits}.

\section{Experiments}\label{sec:experiments}

\subsection{Data collection and organization}

Measurements are performed using \textsc{Labber} \cite{Labber-website}, a commercial software application for data acquisition and instrument control, and \textsc{PSICT} \cite{PSICT-website}, a free and open-source \textsc{Python} module for higher-level scripting.

All measurements presented in the main text and Supplementary Materials were taken during a single cooldown of the dilution refrigerator. 
Two complete datasets related to the magnon detection sensitivity are presented in the main text: the first corresponds to the data taken as a function of sensing time with a nominal Ramsey detuning~\mbox{$\Delta_\mathrm{s}/2\pi=\SI{0}{\MHz}$}~[Fig.\;3] and the second corresponds to the data taken as a function of Ramsey detuning with a fixed sensing time $\tau=\SI{0.8}{\mu\second}$~[Fig.\;4].
The data presented in Fig.\;2 is part of the second dataset, with a nominal Ramsey detuning $\Delta_\mathrm{s}/2\pi = \SI{0}{\MHz}$.

Each of these datasets consists of repeated instances of the measurements to calibrate the magnon detection sensitivity (as described in the main text), interleaved with relatively-rapid measurements of the qubit relaxation and coherence times (as described in Section~\ref{sec:qubit-char}). Additionally, the main datasets are interleaved with qubit-assisted spectroscopy of the Kittel mode (as described in Section~\ref{sec:magnon-spectroscopy}) and characterization of the dispersive shift and calibration of the magnon population (as described in Sections~\ref{sec:strong-dispersive-regime} and \ref{sec:calibration-magnon-population}).

Where applicable, physical quantities quoted in figure captions refer to the values obtained from the specific measurements shown in the figures.
However, simulations or further analysis use averaged values from multiple such measurements, such as the parameters listed in Table~\ref{tab:sim-param-values}.

\subsection{Device overview}

\begin{figure*}[tb]
  \includegraphics[]{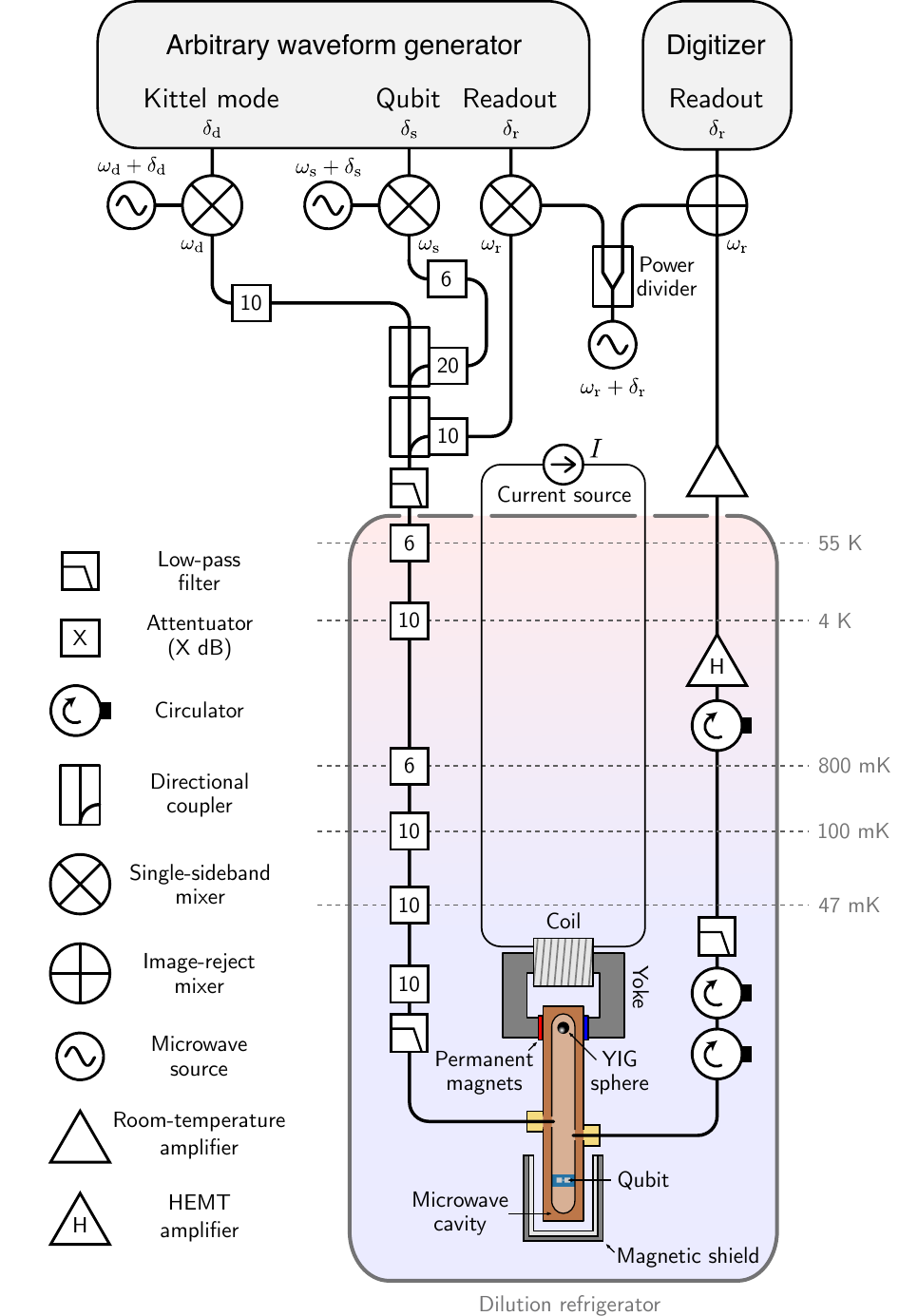}
  \caption{
  Setup used for the experiments, showing the hybrid system within a dilution refrigerator, as well as connections between the microwave control lines and the dc circuit used for controlling the coil current. 
  The base temperature of the refrigerator is $\sim\SI{47}{\milli\kelvin}$.
  The setup is the same as in~Ref.\;\cite{Lachance-Quirion2020}.
  }
  \label{fig:full-exp-setup}
\end{figure*}

The device used in the experiments is the same as that used in Refs.\;\cite{Lachance-Quirion2017,Lachance-Quirion2019,Lachance-Quirion2020}, and the experimental setup is the same as that of Ref.\;\cite{Lachance-Quirion2020}. 
The device is placed within a dilution refrigerator with a base temperature of $\sim\SI{47}{\milli\kelvin}$, where all measurements are performed~[Fig.\;\ref{fig:full-exp-setup}].
The device consists of a microwave copper cavity, a superconducting transmon qubit, and a yttrium-iron-garnet (YIG) sphere. 
The lowest-frequency modes of the cavity are the $\mathrm{TE}_{10p}$ modes, with $p = 1,2,3,4$. 
The parameters of these modes are listed in Table~\ref{tab:params-cavity}.
The parameters of the qubit and Kittel mode of the YIG sphere are listed in Table~\ref{tab:params-other}.

A magnetic circuit, composed of permanent magnets, an iron yoke, and a superconducting coil, is mounted outside the cavity such that the YIG sphere inside the cavity is situated within the generated field. 
The current $I$ in the coil is used for \emph{in situ} tuning of the magnetic field applied to the YIG~sphere, with a conversion ratio of $\SI{1.72}{\milli\tesla\per\milli\ampere}$.
The Kittel mode frequency with the qubit in the ground state $\omega_\mathrm{m}^g$ varies with coil current according to the linear relation
\begin{equation}
  \omega_\mathrm{m}^g(I) = \omega_\mathrm{m}^g(0) + \xi I,
\end{equation}
with \mbox{$\omega_\mathrm{m}^g(0)/2\pi = \SI{8.128}{\GHz}$} and \mbox{$\xi/2\pi = \SI{48.2}{\MHz\per\mA}$}.
In the experiments presented here, the Kittel mode frequency is tuned between the strong dispersive regime at \mbox{$\omega_\mathrm{m}^g/2\pi\approx\SI{7.78}{\GHz}$} ($I=\SI{-8.10}{\milli\ampere}$) and resonance with the $\mathrm{TE}_{102}$ cavity mode at \mbox{$\omega_\mathrm{m}^g/2\pi\approx\SI{8.45}{\GHz}$} ($I=\SI{6.30}{\milli\ampere}$).

\subsection{Preliminary characterization}\label{sec:preliminary-characterization}

\begin{figure*}[tb]
  \includegraphics[]{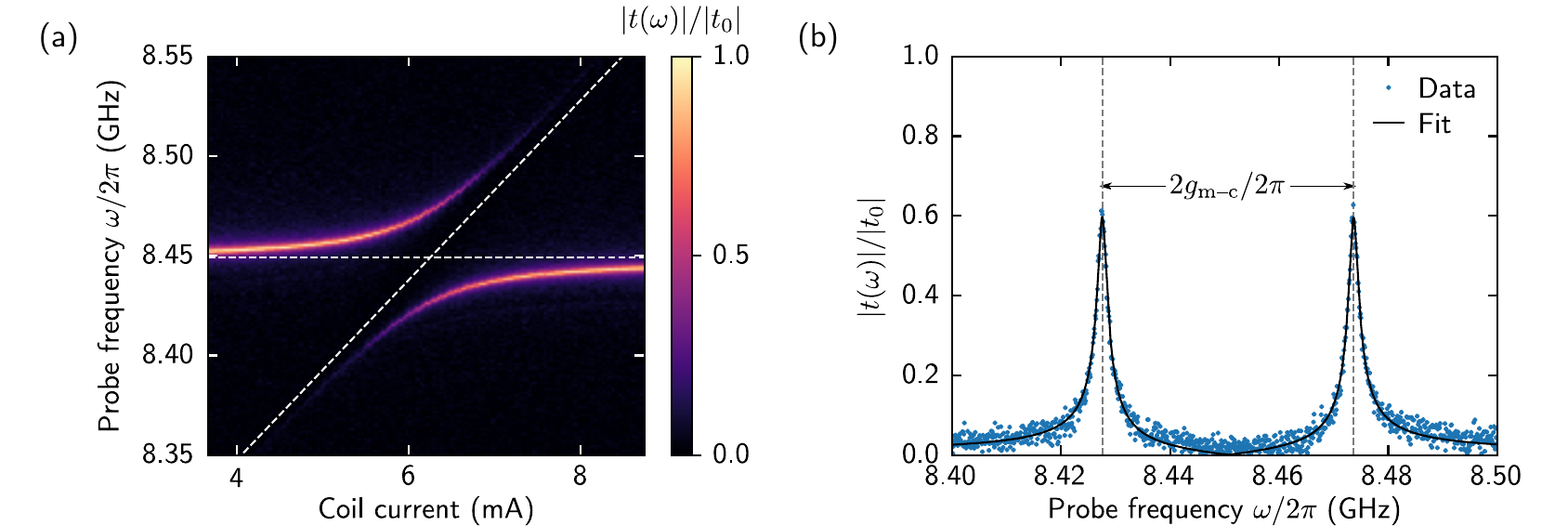}
  \caption{
  Magnon-vacuum-induced Rabi splitting of the $\mathrm{TE}_{102}$ cavity mode resonance.
  (a) Normalized cavity transmission~$\lvert t(\omega)\rvert/\lvert t_0\rvert$ as a function of coil current, showing the avoided crossing between the $\mathrm{TE}_{102}$ cavity mode and the Kittel mode.
  The dashed horizontal and slanted white lines show the frequencies of the $\mathrm{TE}_{102}$ cavity mode and Kittel mode, respectively, bare of their interaction with each other.
  (b) Normalized cavity transmission~$\lvert t(\omega)\rvert/\lvert t_0\rvert$ at $I=\SI{6.31}{\milli\ampere}$, with the $\mathrm{TE}_{102}$ cavity mode and Kittel mode maximally hybridized, and on resonance to within fitting uncertainty.
  The cavity--magnon coupling strength $g_\textrm{m--2}/2\pi=\SI{23.0}{\MHz}$ is half of the size of the splitting between the peaks.
  }
  \label{fig:cavity-magnon-crossing}
\end{figure*}

Preliminary characterization of the interactions and coupling strengths between the systems within the hybrid device is carried out by continuous-wave spectroscopy using a vector network analyzer.
All parameter values described in this section are given in Tables~\ref{tab:params-cavity} and \ref{tab:params-other}.
The qubit--cavity couplings~\mbox{$g_{\textrm{q--}\cavp}$} are not measured directly, but are calculated based on measured parameters. 
In particular,
\begin{equation}
  g_{\textrm{q--}\cavp} 
  = \sqrt{
        \left(\omega_\cavp - \omega_\cavp^g\right)
        \left(\omega_\mathrm{q}^0 - \omega_\cavp - \sum_{p'}\left(\omega_{p'} - \omega_{p'}^g\right)\right)
      },
\end{equation}
where $\omega_\cavp$ is the bare frequency of the $\mathrm{TE}_{10p}$ cavity mode~\cite{Koch2007}.
The Kittel mode--cavity coupling $g_\textrm{m--2}$ is obtained from a measurement of the avoided crossing when the Kittel mode is brought close to resonance with the $\mathrm{TE}_{102}$ mode~[Fig.\;\ref{fig:cavity-magnon-crossing}].
The transmission coefficient $t(\omega)$ can be used to derive the normalized transmission coefficient~\cite{Lachance-Quirion2020}
\begin{equation}\label{eq:t-norm-mc}
  \lvert t(\omega)\rvert/\lvert t_0\rvert = \frac{\kappa_\cavp/2}{\left| i(\omega - \omega_\cavp^g) - \kappa_\cavp/2 + \frac{\lvert g_{\textrm{m--}\cavp}\rvert^2}{i(\omega - \omega_\mathrm{m}^g) - \gamma_\mathrm{m}/2}\right|},
\end{equation}
where $t_0$ is the maximal measured value of the transmission coefficent $t(\omega)$.
Equation~\eqref{eq:t-norm-mc} is fitted to data in Fig.\;\ref{fig:cavity-magnon-crossing}(b) to determine $g_\textrm{m--2}/2\pi=\SI{23}{\MHz}$ and $\gamma_\mathrm{m}/2\pi=\SI{1.48}{\MHz}$, with the $\mathrm{TE}_{102}$ cavity mode and Kittel mode on resonance to within fitting uncertainty.
The cavity linewidth $\kappa_\cavp$ is obtained from measurements with the Kittel mode far detuned from the cavity mode, and is kept fixed when fitting to data.

To carry out continuous-wave spectroscopy of the qubit, the vector network analyzer is set to probe the cavity transmission coefficient at a fixed probe frequency of $\omega/2\pi=\SI{10.4453}{\GHz}$, while a microwave source applies a continuous drive to the device at the spectroscopy frequency $\omega_\mathrm{s}$.
The probe frequency is chosen to maximize the difference in cavity transmission due to the dispersive shift of the $\mathrm{TE}_{103}$ cavity resonance when the qubit is excited, and as such is close to the frequency~$\omega_3$.
The corrected change $\Delta t(\omega_\mathrm{s})$ in the cavity transmission coefficient is obtained by a transformation of the raw signal $t(\omega_\mathrm{s})$ in phase space according to 
\begin{equation}
  \Delta t(\omega_\mathrm{s}) = \mathrm{Re}\left[\mathcal{Z_\text{CW}}(\vartheta)\left(t(\omega_\mathrm{s}) - t_\circ\right)\right],
\end{equation}
where $t_\circ$ is taken to be the centroid of the data in the complex plane, and $\mathcal{Z}_\text{CW}$ is the rotation matrix for the angle $\vartheta=\mathrm{arg}[t_\circ]$.

\begin{figure*}[tb]
  \includegraphics[]{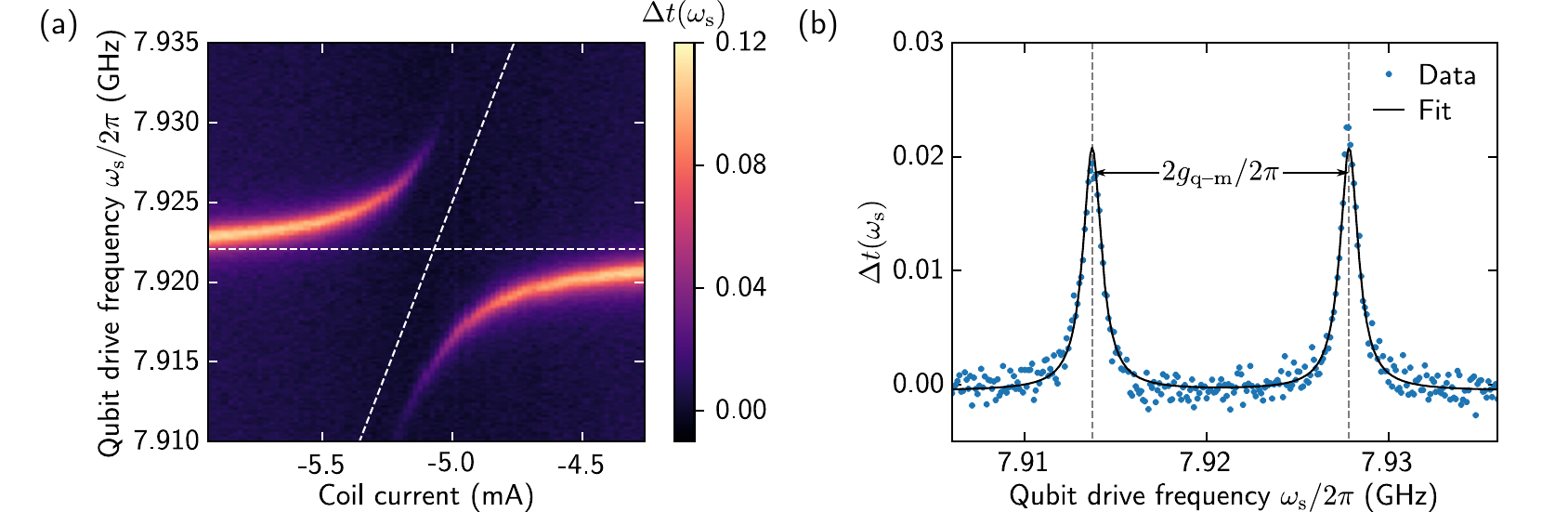}
  \caption{
  Magnon-vacuum-induced Rabi splitting of the qubit transition.
  The qubit spectrum is obtained by measuring the change in transmission coefficient $\Delta t(\omega_\mathrm{s})$ while sweeping the qubit drive frequency $\omega_\mathrm{s}/2\pi$.
  (a) Qubit spectrum as a function of coil current, showing the avoided crossing with the Kittel mode.
  The dashed horizontal and slanted white lines show guides to the eye of the frequencies of the qubit and the Kittel mode, respectively, bare of their interaction with each other.
  (b) Qubit spectrum at $I=\SI{-5.112}{\milli\ampere}$, with the qubit transition and Kittel mode nearly maximally hybridized.
  The qubit--magnon coupling strength $g_\textrm{q--m}/2\pi=\SI{7.07}{\MHz}$ is half of the size of the splitting between the peaks.
  }
  \label{fig:qubit-magnon-crossing}
\end{figure*}

The qubit--Kittel mode coupling $g_\textrm{q--m}$ is based on a second-order cavity-mediated interaction (see Section \ref{sec:theory-governing-equations}). 
Calculating the coupling strength using Eq.\;\eqref{eq:g-q-m} for the lowest four cavity modes gives $g_\textrm{q--m}/2\pi=\SI{7.00}{\MHz}$, while numerically calculating the hybrid system spectrum gives $g_\textrm{q--m}/2\pi=\SI{6.32}{\MHz}$. 
To obtain this value from the experiment, a measurement is taken of the qubit spectrum while the Kittel mode frequency is brought close to resonance with the qubit transition~[Fig.\;\ref{fig:qubit-magnon-crossing}].
When the qubit and Kittel mode are on resonance, the spectrum takes the form of a double Lorentzian
\begin{equation}
  \Delta t(\omega_\mathrm{s}) = 
    C_0 \left[\frac{\gamma_\textrm{q--m}^2/4}{\left(\omega_\mathrm{s} - (\omega_\mathrm{q} - g_\textrm{q--m})\right)^2 + \gamma_\textrm{q--m}^2 / 4} + \frac{\gamma_\textrm{q--m}^2 / 4 }{\left(\omega_\mathrm{s} - (\omega_\mathrm{q} + g_\textrm{q--m})\right)^2 + \gamma_\textrm{q--m}^2 / 4}\right] + \Delta t_0,
\end{equation}
where $C_0$ and $\Delta t_0$ are constants and $\gamma_\textrm{q--m}$ is the average of the qubit and Kittel mode linewidths. 
Fitting this to the data in Fig.\;\ref{fig:qubit-magnon-crossing}(b) gives $g_\textrm{q--m}/2\pi=\SI{7.07}{\MHz}$, showing very good agreement with the theory.

\begin{figure*}[tb]
  \includegraphics[]{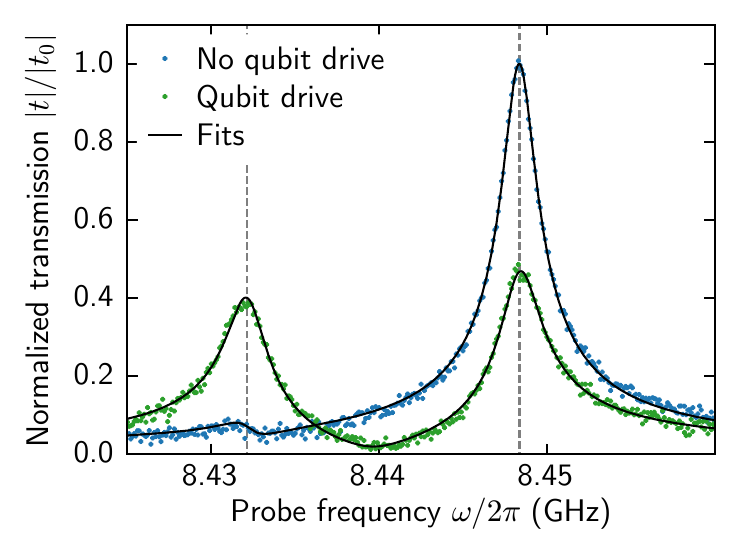}
  \caption{
  Cavity mode transmission spectra with no qubit drive applied (blue) and a strong steady-state qubit drive (green). 
  Gray dashed lines indicate the cavity frequency $\omega_{2}^g/2\pi$ ($\omega_{2}^e/2\pi$) when the qubit is in the ground state (excited state).
  In the case where no qubit drive is applied, the relative amplitude of the peak at the $\omega_{2}^e$ frequency compared to the sum of the amplitudes of the peaks at both frequencies gives the qubit thermal population $p_e^\text{th} = 0.035(6)$.
  }
  \label{fig:qubit-thermal-population}
\end{figure*}

The qubit excited state probability due to thermal excitation, $p_e^\textrm{th}$, is obtained through measurement of the $\mathrm{TE}_{102}$ mode spectrum with and without a deliberate qubit excitation as shown in Fig.\;\ref{fig:qubit-thermal-population}. 
Working within the strong dispersive regime of the cavity--qubit interaction ensures that the dressed peaks $\omega_{2}^g$ and $\omega_{2}^e$, corresponding to the cavity frequency with the qubit in the ground and excited state respectively, are resolved.
Calculating the relative weight of the $\omega_{2}^e$ peak yields $p_e^\textrm{th} = 0.035$.

\subsection{General details for time-resolved measurements}

All time-resolved measurements, discussed in detail in Sections\;\ref{sec:qubit-char} to \ref{sec:calibration-magnon-population}, are carried out using pulse sequences generated by 
an arbitrary waveform generator module, and upconverted to microwave frequencies, as in Fig.\;\ref{fig:full-exp-setup}.
The arbitrary waveform generator module has a resolution of $\SI{1}{\nano\second}$, and the intermediate frequencies used for the qubit readout, qubit control, and magnon excitation are $\delta_\mathrm{r}/2\pi=\SI{90}{\MHz}$, $\delta_\mathrm{s}/2\pi=\SI{95}{\MHz}$, and $\delta_\mathrm{d}/2\pi=\SI{100}{\MHz}$, respectively.
All pulse sequences are followed by a $\SI{400}{\nano\second}$ readout pulse with a square envelope. 
This pulse is transmitted through the device, downconverted, and measured by a signal digitizer with $\SI{2}{\nano\second}$ resolution, with a demodulation window of $\SI{250}{\nano\second}$ selected to optimize the signal-to-noise ratio.
The total sequence duration is $\tau_\textrm{total} = \SI{5}{\mu\second}$, which results in a repetition rate of $\SI{0.2}{\MHz}$.
The chosen value was selected based on measuring the sensitivity as a function of $\tau_\textrm{total}$, with an optimum resulting from the competing effects of measurement-induced dephasing due to a residual population of cavity photons, and the increase in SNR resulting from a higher repetition rate.
This is much longer than the qubit relaxation time $T_1$, \mbox{$\tau_\textrm{total} \gg T_1 \approx \SI{0.80}{\mu\second}$}, so the qubit is initialized to the ground state~$\ket{g}$ before each shot through relaxation.
Table~\ref{tab:number-of-shots} summarizes the number of shots used when repeating sequences for different measurements.

\begin{table*}[tb]
  \caption{Number of shots used in different experiments described in the main text and supplementary material. The statistical error is given by $1/\sqrt{N}$.}
  \begin{ruledtabular}
  \begin{tabular*}{\textwidth}{l@{\extracolsep{\fill}}ccd{5.5}}
    Measurement & Figure & Number of shots $N$ & \multicolumn{1}{c}{Statistical error (\%)} \\
    \hline
    Qubit spectrum with magnon-number splitting & 1(d) & $\num{5e5}$ & 0.14 \\
    Qubit readout calibration & \ref{fig:qubit-readout} & $\num{1e5}$ & 0.32 \\
    Magnon detection signal measurement & 2(a) & $\num{1e6}$ & 0.1 \\
    Qubit noise measurement & 2(b), 3(b) & $\num{1e6}$ & 0.1 \\
    \hline
    Qubit Ramsey & \ref{fig:qubit-char}(b) & $\num{4e4}$ & 0.5 \\
    Qubit $T_1$ & \ref{fig:qubit-char}(a) & $\num{1e4}$ & 1.0 \\
    Qubit-assisted Kittel mode spectroscopy & \ref{fig:magnon-spectroscopy} & $\num{1e4}$ & 1.0 \\
  \end{tabular*}
  \end{ruledtabular}
  \label{tab:number-of-shots}
\end{table*}

\subsection{Qubit characterization}\label{sec:qubit-char}

\begin{figure*}[tb]
  \includegraphics[]{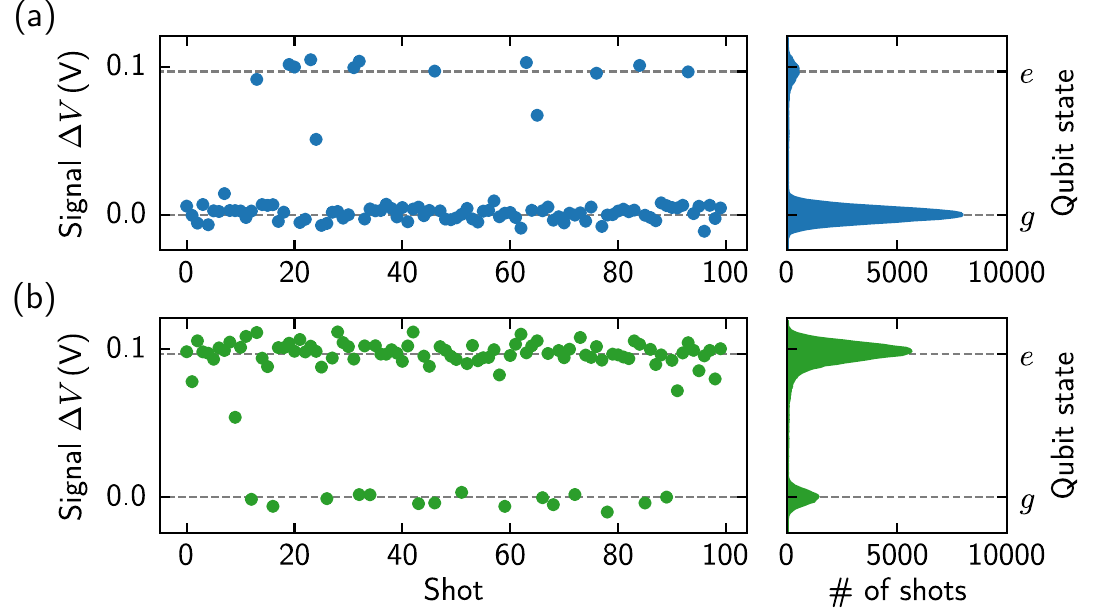}
  \caption{
  High-power readout of qubit when preparing (a) the ground state $\ket{g}$ and (b) the excited state $\ket{e}$, showing samples of demodulated signal voltage $\Delta V$ (left) and histograms of the complete datasets (right).
  Gray dashed lines are the signal values identified as corresponding to the ground and excited states of the qubit.
  }
  \label{fig:qubit-readout}
\end{figure*}

The qubit is measured using the high-power readout technique \cite{Reed2010}, following the procedure described in Ref.\;\cite{Lachance-Quirion2020}. 
In particular, the corrected demodulated signal $\Delta V$ is obtained by a transformation of the raw demodulated signal, $V$, in phase space according to
\begin{equation}
  \Delta V = \mathrm{Re}\left[\mathcal{Z}(\theta)\left(V - V_g\right)\right],
\end{equation}
where $\mathcal{Z}(\theta)$ is the rotation matrix for an angle
\begin{equation}
  \theta = \arctan\left(\frac{\mathrm{Im}[V_e] - \mathrm{Im}[V_g]}{\mathrm{Re}[V_e] - \mathrm{Re}[V_g]}\right),
\end{equation}
and $V_g$ and $V_e$ are the raw demodulated signals corresponding to the qubit in the ground state and excited state, respectively.
The signal $\Delta V$ can then be mapped to the qubit state as, by construction, $\Delta V = 0$~($\Delta V = \lvert V_e-V_g\rvert$) corresponds to $p_e = 0$~($p_e = 1$). 
Values of $\Delta V$ that fall in between $V_g$ and $V_e$ correspond to $p_e$ given by
\begin{equation}
  p_e = \frac{1}{N}\frac{\sum_{n=1}^{N}\mathrm{Re}\left[\mathcal{Z}(\theta)V_n - V_g\right]}{\mathrm{Re}\left[\mathcal{Z}(\theta)V_e - V_g\right]},
\end{equation}
where $V_n$ is the demodulated signal of shot $n$, and $N$ is the total number of shots~[Fig.\;\ref{fig:qubit-readout}].
The qubit is excited using Gaussian-shaped pulses with an envelope
\begin{equation}\label{eq:qubit-pulse-envelope}
  \Omega_\mathrm{s}(t)  = \Omega_\mathrm{s} e^{-\pi(t-t_\mathrm{s})^2/\tau_\mathrm{s}^2},
\end{equation}
where $t_\mathrm{s}$ is the pulse center time, and $\tau_\mathrm{s}$ is pulse width. 
The qubit $\pi$-pulse amplitude $A_\pi$ is calibrated for fixed pulse widths $\tau_\mathrm{s}$ of $\SI{200}{\nano\second}$ (qubit-assisted Kittel mode spectroscopy) and $\SI{12}{\nano\second}$ (all other sequences) by fitting the demodulated signal, $\Delta V(A_\mathrm{s})$, obtained for different pulse amplitudes to
\begin{equation}
  \Delta V(A_\mathrm{s}) = \frac{\Delta V_e}{2}\left[ 1 - \cos\left( \pi\frac{A_\mathrm{s}}{A_\pi} \right) \right],
\end{equation}
where $A_\mathrm{s}$ is the pulse amplitude used, and $\Delta V_e$ is the demodulated signal corresponding to the qubit in the excited state.

\begin{figure*}[tb]
  \includegraphics[]{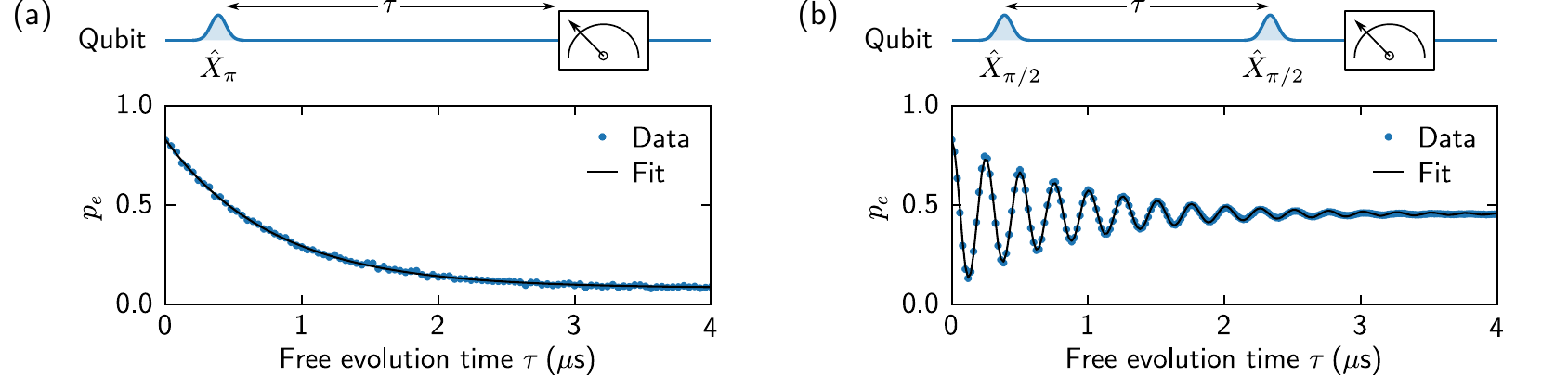}
  \caption{
  (a) Pulse sequence and results of characterizing the relaxation time~$T_1$ of the qubit.
  An $\hat{X}_\pi$ pulse is applied to the qubit to populate the excited state~$\ket{e}$, after which the qubit is allowed to relax for a time~$\tau$ before measurement.
  The solid black line is a fit to a decaying exponential.
  (b) Pulse sequence and results of characterizing the coherence time~$T_2^\ast$ and the frequency $\omega_\mathrm{q}$ of the qubit through Ramsey interferometry.
  Two $\hat{X}_{\pi/2}$ pulses, separated by a free evolution time~$\tau$, are applied to the qubit, inducing Ramsey oscillations as the sequence is repeated for different values of $\tau$.
  The solid black line is a fit to Eq.\;\eqref{eq:pe-ramsey-plain}.
  }
  \label{fig:qubit-char}
\end{figure*}

The qubit relaxation time $T_1\approx \SI{0.80}{\mu\second}$ is measured by exciting the qubit with a $\pi$-pulse, and allowing it to relax for a free evolution time $\tau$ as shown in Fig.\;\ref{fig:qubit-char}(a). The relaxation time $T_1$ is obtained as the time constant of a decaying exponential fitted to the data. 
The relaxation time is limited by Purcell decay through the cavity modes. In particular, the theoretical maximum can be calculated by considering~\cite{Koch2007}
\begin{equation}\label{eq:gamma-q-1}
  \gamma_{1} = \sum_p \kappa_\cavp \left(\frac{g_{\textrm{q--}\cavp}}{\omega_\cavp-\omega_\mathrm{q}}\right)^2,
\end{equation}
with $T_1 = 1/\gamma_{1}$. Using values for the lowest three cavity modes, the Purcell limit of the relaxation time is $\textrm{max}[T_1] = \SI{0.88}{\mu\second}$.
The qubit coherence time $T_2^\ast \approx \SI{0.89}{\mu\second}$ is measured using Ramsey interferometry as shown in Fig.\;\ref{fig:qubit-char}. 
Two $\pi/2$ pulses of amplitude $A_\pi/2$ are applied to the qubit, separated by a peak-to-peak free evolution time $\tau$ and detuned by \mbox{$\Delta_\mathrm{s}/2\pi=\SI{-4}{\MHz}$}. 
The coherence time $T_2^\ast$ is extracted by fitting the data to Eq.\;\eqref{eq:pe-ramsey-plain}.
The upper limit of the coherence time is given by double the relaxation time, that is, $\textrm{max}[T_2^\ast] = 2T_1\approx\SI{1.60}{\mu\second}$.
The difference between the upper limit and the value observed in the experiments is probably due to pure dephasing from the thermal population of the cavity modes.
Assuming the lowest three cavity modes have an equal thermal population, it would need to be around $\overline{n}_\mathrm{c}^\text{th}\approx 0.025$ to entirely account for this discrepancy.

The qubit readout errors are obtained as in Ref.\;\cite{Lachance-Quirion2020}. For the results presented in the main text, numerical simulations assuming no qubit readout errors improve the resulting sensitivity by around $10\%$, again showing that the qubit readout is not the dominant limiting factor with respect to the magnon detection sensitivity.

\subsection{Qubit-assisted Kittel mode spectroscopy}\label{sec:magnon-spectroscopy}

\begin{figure*}[tb]
  \includegraphics{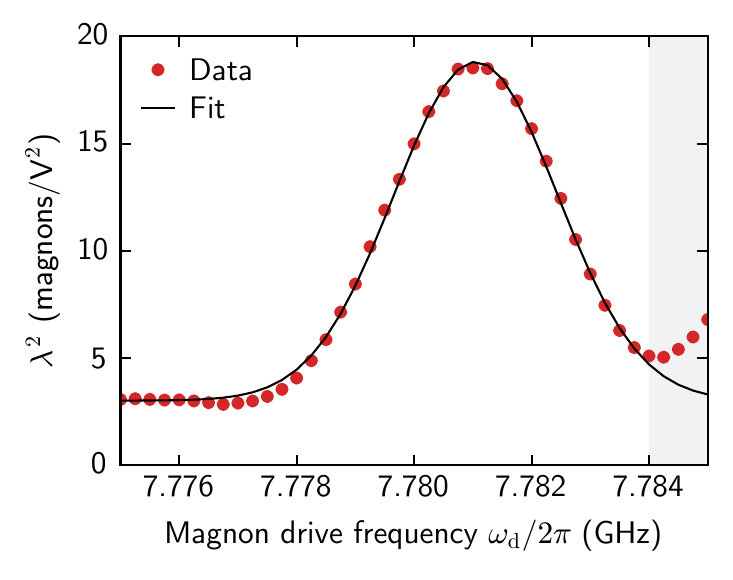}
  \caption{
  Indirect qubit-assisted spectroscopy of the Kittel mode, presented as the squared coefficient $\lambda^2$ as a function of the pump frequency $\omega_\mathrm{d}/2\pi$ at which the displacement pulse is applied. Data is taken at a coil current $I = \SI{-8.10}{\milli\ampere}$. The solid black line shows a fit to a Gaussian function with a vertical offset. The data in the shaded area is not considered for fitting due to close proximity to the $\ket{e}\leftrightarrow\ket{f}$ qubit transition.
  The dataset shown in this figure gives a magnon frequency of \mbox{$\omega_\mathrm{m}^g/2\pi = \SI{7.78105(2)}{\GHz}$}.
  }
  \label{fig:magnon-spectroscopy}
\end{figure*}

Spectroscopy of the Kittel mode is carried out indirectly via the methodology described in~Ref.\;\cite{Lachance-Quirion2020}. 
The procedure involves preparing a coherent state of magnons in the Kittel mode with a displacement pulse with a maximal amplitude~$A_\mathrm{d}$ applied at a frequency~$\omega_\mathrm{d}$.
Following this, a qubit excitation conditional on the magnon population being zero is carried out. 
For a given value of $A_\mathrm{d}$, the resulting magnon population $\overline{n}_\mathrm{m}$ is larger as the detuning from the Kittel mode decreases. The coefficient $\lambda$ can be measured, where
\begin{equation}\label{eq:magnon-lambda}
  \overline{n}_\mathrm{m} = \left( \lambda A_\mathrm{d} \right)^2.
\end{equation}
The variation of $\lambda^2$ with respect to the magnon pump frequency $\omega_\mathrm{d}$ represents the magnon spectrum, as shown in Fig.\;\ref{fig:magnon-spectroscopy}.
In the experiments presented in this Letter, the coil current is fixed at $I = \SI{-8.10}{\milli\ampere}$. 
The dressed Kittel mode frequency is, on average, $\omega_\mathrm{m}^g/2\pi = \SI{7.78106}{\GHz}$. 
In the experiments, there is an unintentional detuning between the magnon drive and the Kittel mode, on average~$\Delta_\mathrm{d}/2\pi=\SI{-42(8)}{\kHz}$, which is much smaller than the magnon linewidth and therefore does not significantly affect the results.

\subsection{Characterization of the strong dispersive regime}\label{sec:strong-dispersive-regime}

\begin{figure*}[tb]
  \includegraphics[]{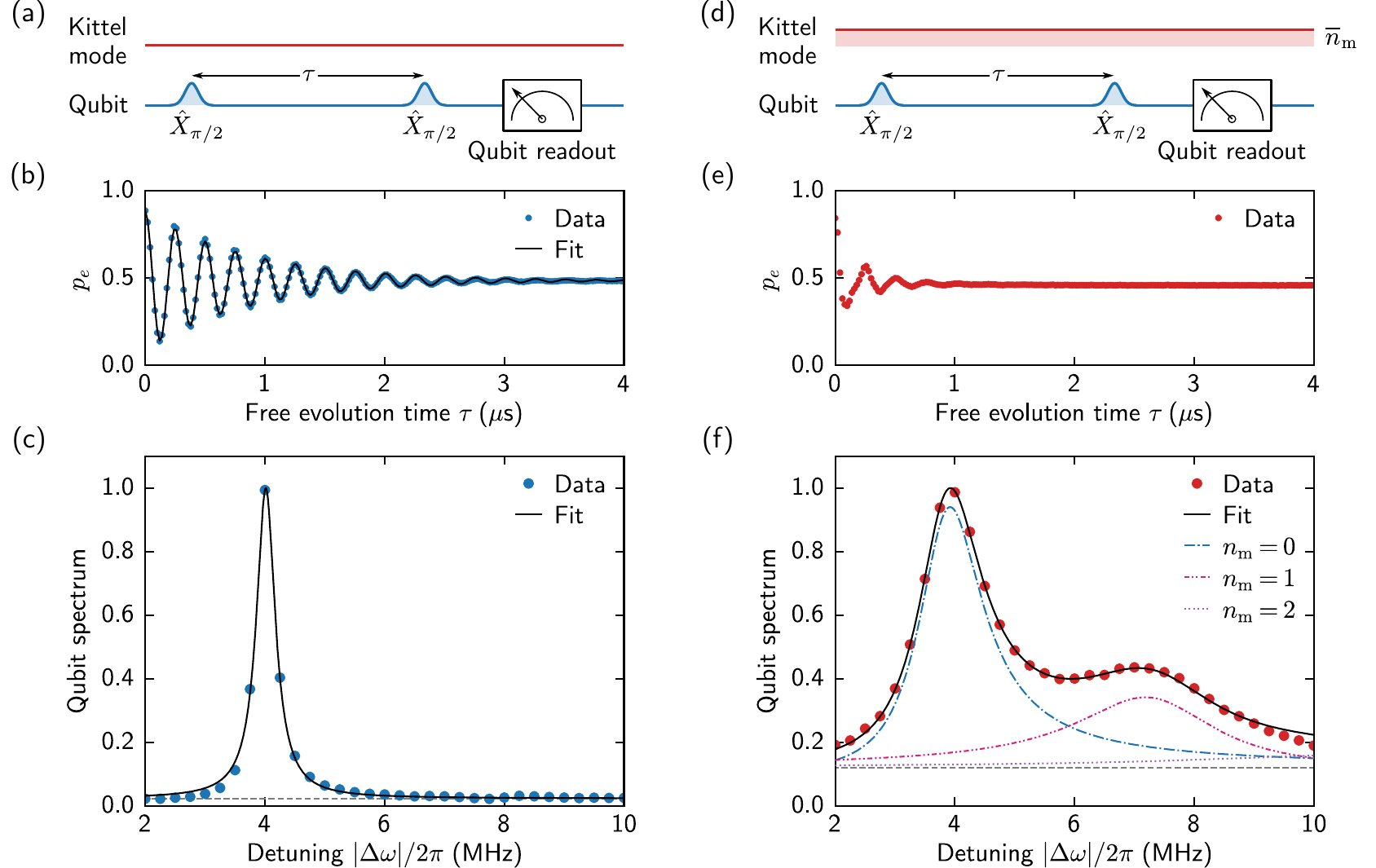}
  \caption{
  (a) Pulse sequence used for observing Ramsey oscillations of the qubit with no magnon population.
  The $\hat{X}_{\pi/2}$ pulses are intentionally detuned from the qubit frequency by $\Delta_\mathrm{s}/2\pi=\SI{-4}{\MHz}$.
  (b) Ramsey interferometry of the qubit, showing the qubit excited state probability~$p_e$ after applying the sequence in (a), as a function of free evolution time~$\tau$.
  The solid line shows a fit to Eq.\;\eqref{eq:pe-ramsey-plain}.
  (c) Qubit spectrum from a Fourier transform of~(b), showing the qubit resonance without magnons in the Kittel mode.
  The solid line is a fit to a Lorentzian function, as in Eq.\;\eqref{eq:qubit-spectrum-magnon-splitting-full} only considering $n_\mathrm{m}=0$.
  The data is normalized to the maximal value of the fit.
  (d) Pulse sequence used for observing Ramsey oscillations of the qubit with a continuous drive applied near resonance with the Kittel mode to excite a magnon population $\nm=0.615$.
  (e) Ramsey interferometry results after applying the pulse sequence in~(d), with the Ramsey oscillations showing that the qubit coherence time is significantly reduced compared to~(b), due to the presence of magnons in the Kittel mode.
  (f) The spectrum obtained from a Fourier transform of~(e), showing the magnon-number splitting of the qubit spectrum.
  The qubit peaks corresponding to the different magnon Fock states $n_\mathrm{m}$ are resolved due to the qubit and Kittel mode being coupled in the strong dispersive regime, where the dispersive shift per excitation $2\lvert\chi_\textrm{q--m}\rvert$, is larger than both the qubit linewidth $\gq$ and magnon linewidth $\gm$.
  The lines show fitting to the models as described in Section~\ref{sec:strong-dispersive-regime}, and the data is normalized to the maximal value of the fit.
  }
  \label{fig:magnon-number-splitting-exp}
\end{figure*}

Measurement of the qubit--magnon dispersive shift $\chi_\textrm{q--m}$ is carried out using the same sequence as is used for characterization of the magnon detection sensitivity~[Fig.\;\ref{fig:magnon-number-splitting-exp}(a) and (d)]. 
In this case, the Kittel mode drive amplitude is larger, in order to obtain a more prominent qubit peak for $n_\mathrm{m}=1$ and thereby ensure robust fitting to the spectrum. 
Carrying out the Ramsey sequence with an intentional detuning $\Delta_\mathrm{s}^0 \equiv \omega_\mathrm{q}^0 - \omega_\mathrm{s} \gg \gamma_\mathrm{q}^0$ leads to oscillations in $p_e(\tau)$~[Fig.\;\ref{fig:magnon-number-splitting-exp}(b) and (e)]. 
The qubit spectrum $S(\Delta\omega)$ is obtained by applying a Fourier transform to $p_e(\tau)$, with
\begin{equation}
  S(\Delta\omega) = \frac{\mathrm{Re}\left[\mathcal{F}\left\{p_e(\tau)\right\}(\Delta\omega) \right]}
                    {\text{max}\left[ \mathrm{Re}\left[\mathcal{F}\left\{p_e(\tau)\right\}(\Delta\omega) \right] \right]}.
\end{equation}

In the absence of a magnon population, the qubit spectrum fits a Lorentzian lineshape as in Fig.\;\ref{fig:magnon-number-splitting-exp}(c), from which a qubit linewidth $\gamma_\mathrm{q}^0/2\pi = \SI{0.377(7)}{\MHz}$ is obtained. 
When a nonzero magnon population is present, the spectrum~$S(\Delta\omega)$ can be described using the model in Section~\ref{sec:theory-strong-dispersive-regime}, in particular with 
\begin{equation}\label{eq:qubit-spectrum-magnon-splitting-full}
  S(\Delta\omega) = C \sum_{n_\mathrm{m}=0}^\infty s_{n_\mathrm{m}}(\Delta\omega) + S_0,
\end{equation}
where $C, S_0$ are free parameters and $s_{n_\mathrm{m}}(\Delta\omega)$ is given by Eq.\;\eqref{eq:qubit-spectrum-model}. 
The model is fitted with a truncated Fock space up to $n_\mathrm{m}=9$, with the qubit linewidth $\gamma_\mathrm{q}^0/2\pi=\SI{0.377}{\MHz}$ and the magnon drive detuning $\Delta_\mathrm{d}/2\pi = \SI{-42}{\kHz}$ fixed. 
Carrying out this fitting on the measurements interleaved with the main datasets yields averaged values of the dispersive shift~\mbox{$\chi_\textrm{q--m}/2\pi=\SI{-1.76}{\MHz}$} and magnon linewidth~\mbox{$\gamma_\mathrm{m}/2\pi=\SI{1.57}{\MHz}$}~[Fig.\;\ref{fig:magnon-number-splitting-exp}(f)].
Additionally, Fig.\;\ref{fig:magnon-number-splitting-exp}(f) also shows the decomposition of the full spectrum into the qubit peaks corresponding to specific magnon Fock states up to $n_\mathrm{m}=2$, by considering only the corresponding $s_{n_\mathrm{n}}(\Delta\omega)$ term in Eq.\;\eqref{eq:qubit-spectrum-magnon-splitting-full}. 

The data and fits shown in Fig.\;1(d) in the main text are the same as those in Fig.\;\ref{fig:magnon-number-splitting-exp}(c) and (f), but with zero values added to the end of the data as a function of free evolution time, in order for the resulting Fourier transform to represent more points on the frequency axis.

\subsection{Calibration of excited magnon population}\label{sec:calibration-magnon-population}

As described in the main text, the results presented in this Letter involve a characterization of the magnon detection sensitivity using a magnon population intentionally excited by a drive resonant or near-resonant with the Kittel mode.
Fitting of the qubit spectrum obtained in the presence of the magnon drive, as described in the previous section, reveals that a pump amplitude $A_\mathrm{d} = \SI{25}{\milli\volt}$ yields a magnon population $\overline{n}_\mathrm{m} = 0.615(12)$~[Fig.\;\ref{fig:magnon-number-splitting-exp}(f)]. 
The coefficient $\lambda$, calculated according to Eq.\;\eqref{eq:magnon-lambda}, is used in order to determine the magnon population generated by a given drive amplitude.


\bibliographystyle{apsrev4-1}

%